\documentclass{raa}
\usepackage{mathrsfs}
\usepackage{graphicx,times}
\usepackage{natbib}
\usepackage{amssymb,amsmath}

\usepackage[pagebackref=true]{hyperref}

\begin{document}

   \title{Formation and Destiny of White Dwarf and Be Star Binaries}

 \volnopage{ {\bf 20XX} Vol.\ {\bf X} No. {\bf XX}, 000--000}
   \setcounter{page}{1}

   \author{ChunHua Zhu\inst{1}, GuoLiang L{\"u}\inst{2,1},  Xizhen Lu\inst{1}, Jie He\inst{1}
   }

   \institute{School of Physical Science and Technology, Xinjiang University, Urumqi 830046, China; {\it chunhuazhu@sina.cn}\\
        \and
            Xinjiang Astronomical Observatory, Chinese Academy of Sciences, 150 Science 1-Street, Urumqi, Xinjiang 830011, China\\
\vs \no
   {\small Received 20XX Month Day; accepted 20XX Month Day}}

\abstract{The binary systems consisting of a Be star and a white dwarf (BeWDs) are very interesting.
They can originate from the binaries composed of a Be star and a subdwarf O or B star (BesdOBs),
and they can merge into red giants via luminous red nova or can evolve into double WD potentially detected
by $LISA$ mission. Using the method of population synthesis,
we investigate the formation and the destiny of BeWDs,
and discuss the effects of the metallicity ($Z$) and the common envelope evolution parameters.
We find that BesdOBs are significant
progenitors of BeWDs. About 30\% ($Z=0.0001$)-50\% ($Z=0.02$) of BeWDs come from
BesdOBs.  About 60\% ($Z=0.0001$) -70\% ($Z=0.02$) of BeWDs turn into red giants via a merger between a WD and a non-degenerated star.
About 30\% ($Z=0.0001$) -40\% ($Z=0.02$) of BeWDs evolve into double WDs which are potential gravitational waves of $LISA$ mission
at a frequency band between about $3\times10^{-3}$ and $3\times10^{-2}$ Hz.
The common envelope evolution parameter introduces
an uncertainty with a factor of about 1.3 on BeWD populations in our simulations.
}

   \authorrunning{C.-H. Zhu et al. }            
   \titlerunning{Formation and Destiny of BeWDs}  
   \maketitle

%
\section{INTRODUCTION}
High mass X-ray binaries (HMXBs) consist of a massive
star and a compact object, in which the massive star may be a red supergiant star or a Be star,
and the compact object may be a black hole or a neutron star (NS).
There are about more than 240 HMXBs observed in the Galaxy and the Magellanic Clouds (MCs) \citep{Liu2005,Liu2006}.
The majority of the known HMXBs are composed of Be stars and NSs (BeNSs), which are called as Be/X-ray binaries.
\cite{Meurs1989} estimated that there should be 2000-20000 Be/X-ray binaries in the Galaxy.

The compact stars in Be/X-ray binaries can also be white dwarfs (WDs). They are marked as BeWDs in this paper.
Based on the model of binary evolution, \cite{Raguzova2001} predicted that the number of BeWDs in the
Galaxy should be 7 times more than that of BeNSs.
Unfortunately, there are only 7 BeWDs or candidates observed in the MCs, which are listed in Table \ref{tab:BeWD}.
Compared with BeNSs, the X-ray spectrum produced by BeWDs is very soft, which is absorbed more easily. Especially,
due to the much higher extinction rates in the plane of the Milky Way, up to now, no BeWD is known in the Galaxy.
Beside the above observational biases, \cite{Kennea2021} considered that the metallicity may significantly
affect the evolution of BeWDs.

In spite of observational constraints, there should be a large number of BeWDs in the Universe. They involve
a Be star and an accreting WD. The former has a B-type spectrum, a high rotational velocity and a decretion disk \citep{Porter2003}.
The origin of this disk is still unclear. The majority of Be stars are usually produced by
binary interaction \citep[e. g.,][]{Ablimit2013,deMink2013,Hastings2021}. Therefore, the progenitor of WD in BeWD
transfers enough matter to spin up its companion, even it can lose whole envelope, and becomes a naked helium star.
On observations, naked helium star is considered as subdwarf O or B (sdOB) star\citep{Sargent1968,Heber1986,Han2002}.
A binary system, consisting a sdOB star and a Be star, is called as BesdOB.
\cite{Naze2022} listed known 25 BesdOB and candidates \citep[][references therein]{Wang2018,Wang2021},
 which are listed in Table \ref{tab:Besd}.
 Obviously, it is very interesting to discuss whether these BesdOBs
can evolve into BeWDs.  The latter is the potential progenitor for type Ia supernova (SN Ia) \citep[e. g.,][]{Wang2012}
or millisecond pulsar (MSP) \citep[e. g.,][]{DAntona2020},
which depends on accreting compact object being CO WD or ONe WD.
In addition, the mass of Be stars is between about 2 and 20 M$_\odot$ \citep{Porter2003}. They finally become WDs or NSs.
Then, BeWDs evolve into systems consisting of double compact objects which are good gravitational sources detected by
the Laser Interferometer Space Antenna ($LISA$) \citep{Amaro-Seoane2017,Lu2020}.

There are many theoretical studies for Be binaries, such as Be/X-ray binaries, BesdOBs, BeWDs,
and so on \citep[e. g.,][]{Brown2019,Raguzova2001,Shao2014,Shao2021}.
Especially, \cite{Brown2019} considered the interaction between NS and decretion disk in Be/X-ray binaries.
However, this interaction seldom is included in BeWDs.
In the present paper, we focus on the formation of BeWDs via BesdOBs or other channels, and their destiny (SN Ia or MSP) when
WD's mass reaches Chandrasekhar mass by accreting the decretion disk of Be star.
In \S\  2,  the assumptions and some details of the modelling
algorithm are given. In \S\  3, the properties of the
model population of BeWDs, their formation channel and destiny are presented. Conclusions follow in \S\  4.

\begin{table*}
  \caption{Parameters of the observed BeWDs. Columns 1 to
       7 list the name of BeWDs, orbital period $P_{\rm orb}$, X-ray luminosity, estimated WD's mass, the spectral type of Be star,
       hosted galaxy and references.   References: K06-\citet{Kahabka2006}; S12-\citet{Sturm2012}; L12-\citet{Li2012};
       M13-\citet{Morii2013};K21-\citet{Kennea2021}; C20-\citet{Coe2020}; O10-\citet{Oliveira2010}.
 }
  \tabcolsep1.0mm
  \begin{tabular}{lcccccc}
  \hline
  BeWD&$P_{\rm orb}$ (days)&$L_{\rm X}$ (erg s$^{-1}$)&$M_{\rm WD}$ (M$_{\odot}$)&Be star&galaxy&References\\
  \hline
 XMMU J052016.0-692505&510 or 1020&10$^{34}$-10$^{38}$&0.9-1.0&B0-B3e&LMC&K06\\
 XMMU J010147-715550&1264&$\sim$4.4$\times$10$^{33}$&1.0&O7IIIe-B0Ie&SMC&S12\\
 MAXI J0158-744&-&$>10^{37}$; $10^{40}$ of peak luminosity&1.35&B1-2IIIe&SMC&L12,M13\\
 SWIFT J011511.0-725611&17.402&2$\times10^{33}$-3.3$\times10^{36}$&1.2&O9IIIe&SMC&K21\\
 SWIFT J004427.3734801&21.5&$5.7-2.9\times10^{36}$&-&O9Ve-B2IIIe&SMC&C20\\
 RX J0527.8-6954&-&$4-9\times10^{36}$&-&B5eV&LMC&O10\\

\hline
 \label{tab:BeWD}
\end{tabular}
\end{table*}

\begin{table*}
  \caption{Parameters of the observed BesdOBs in the Galaxy. Columns 1 to
       9 list the name of BesdOB, orbital period $P_{\rm orb}$, mass, effective temperature and luminosity of sdOB,
       mass, effective temperature and luminosity  of Be star and references.
       References: K22-\citet{Klement2022}; K12-\citet{Koubsky2012}; M15-\citet{Mourard2015};
       P13-\citet{Peters2013}; P16-\citet{Peters2016}; C18-\citet{Chojnowski2018};B20-\citet{Bodensteiner2020};
       R09-\citet{Ruzdjak2009}; S20-\citet{Shenar2020}; C02-\citet{Carrier2002}; A78-\citet{Abt1978}; H87-\citet{Harmanec1987};
       T08-\citet{Tycner2008}; K14-\citet{Koubsky2014}; W21-\citet{Wang2021}.
 }
  \tabcolsep1.0mm
  \begin{tabular}{ccccccccc}
  \hline
  BesdOB&$P_{\rm orb}$ (days)&$M_{\rm sdOB}$ (M$_{\odot}$)&$T_{\rm eff}$ (kK)(sdOB)&$\log L (L_\odot)$ (sdOB)&$M_{\rm Be}$ (M$_{\odot}$)&$T_{\rm eff}$ (kK)(Be)&$\log L (L_\odot)$ (Be)&References\\
  \hline
  V2119 Cyg&63.1&$1.62\pm0.28$&43.5&$2.92^{0.15}_{-0.23}$&$8.65\pm0.35$&25.6&$3.83\pm0.02$&K22\\
  60 Cyg&147.68&$1.2\pm0.2$&42&2.78&$7.3\pm1.1$&27&$3.99\pm0.04$&K22\\
  28 Cyg&246&-&45&$<$2.39&-&20.47&$3.76\pm0.02$&K22\\
  $o$ Puppis&28.9&-&-&-&-&-&-&K12\\
  $\varphi$ Persei&126.67&$1.2\pm0.2$&53&$3.79\pm0.13$&$9.6\pm0.3$&29.3&$4.16\pm0.04$&M15\\
  HR 2142&80.9&0.7&$>43$&$>1.7$&9&21&$4.17\pm0.10$&P16\\
  59 Cygni&28.2&0.79&52.1&$3.0\pm0.1$&7.9&21.8&$4.14\pm0.12$&P13\\
  FY CMa&37.3&1.26&45&3.38&12.6&27.5&$4.43\pm0.03$&P13\\
  HD 55606&93.8&0.9&40.9&$2.27^{+0.13}_{-0.19}$&6.2&27.35&$3.60\pm0.03$&C18\\
  HR 6819 & 40.335 & $0.46\pm0.26$&16&$3.12\pm0.10$&$7\pm2$&20&$3.77\pm0.04$&B20\\
  $\zeta$ Tau & 132.987 & 0.87-1.02&-&-&11&19.3&$3.75\pm0.04$&R09\\
  AlS8775 & 78.999 &1.5&12.7&2.8&$7\pm2$&18&$3.10\pm0.07$&S20\\
  MX Pup & 5.1526 &0.6-6.6&-&-&15&25.1&$4.24\pm0.03$&C02\\
  $\chi$ Oph &34.1 or 138.8 &3.8&-&-&10.9&20.9&$3.75\pm0.02$&A78, H87,T08\\
  HD 161306&99.9&0.0567&-&-&-&-&$3.56\pm0.01$&K14\\
  V1150 Tau & - &-&40&$2.11^{\rm+0.14}_{-0.21}$&-&20.53&$3.47\pm0.02$&W21\\
  HR2249 & - &-&38.2&$2.68^{\rm+0.15}_{-0.23}$&8.5&21.5&$3.55\pm0.02$&W21\\
  QY Gem & - &-&43.5&$2.75^{\rm+0.13}_{-0.18}$&-&20&$3.49\pm0.03$&W21\\
  V378 Pup & - &-&42&$2.83^{\rm+0.14}_{-0.20}$&-&20&$3.99\pm0.03$&W21\\
  LS Mus & - &-&45&$2.82^{\rm+0.23}_{-0.53}$&-&22.8&$3.86\pm0.03$&W21\\
  kap01 Aps & - &-&40&$2.64^{\rm+0.14}_{-0.20}$&-&23.95&$3.83\pm0.02$&W21\\
  V846 Ara & - &-&42&$2.28^{\rm+0.14}_{-0.21}$&-&19.8&$3.39\pm0.01$&W21\\
  $\iota$ Ara & - &-&33.8&$2.60^{\rm+0.15}_{-0.23}$&-&25.86&$3.95\pm0.02$&W21\\
  V750 Ara & - &-&45&$<2.61$&-&25&$4.49\pm0.04$&W21\\
  8Lac A & - &-&45&$<2.71$&-&27.38&$4.17\pm0.05$&W21\\

\hline
 \label{tab:Besd}
\end{tabular}
\end{table*}

\section{Model}
BeWDs' formation and evolutionary involve binary interactions: tidal interaction, mass transfer, common envelope evolution (CEE),
and so on.  We use the rapid binary
star evolution (BSE) code \citep{Hurley2002,Kiel2006}.
In BSE code, the stellar structure and evolution is described
by a serial of fitting formula which depend on the stellar mass, the metallicity and evolutionary age \citep{Hurley2000}.
The binary evolution is determined by some binary interactions: mass transfer, tidal interaction,
CEE, gravitational radiation, magnetic braking, coalescence.
Among them, the mass transfer, the tidal interaction and the magnetic braking can
directly impact on the stellar rotation.
These interactions may introduce some uncertain input parameters.
If any input parameter is not specially
mentioned in the next subsection, it is taken as the default value in these papers.

\subsection{Be star}
The mean value of rotational velocities of Be stars is about $70\%-90\%$ of the Keplerian  critical velocity ($v_{\rm cr}$),
and the range of their masses is between 3 and 22 M$_\odot$ \citep{Ekstrom2008,Porter2003}.
Following \cite{Ekstrom2008}, we assume that a main-sequence star becomes Be star when its rotational velocity exceeds 0.7 $v_{\rm cr}$ and
its mass is  between 3 and 22 M$_\odot$.

The formation of Be star has been investigated by a lot of works \citep[e. g.,][]{Ekstrom2008,deMink2013,Shao2014,Shao2021}.
They meticulously discussed the effects of many uncertain parameters, which include the critical mass ratio ($q_{\rm cr}$) for dynamically unstable mass transfer,
the efficiency of mass accretion when the accretor is spun up to $v_{\rm cr}$,
the combining parameter $\lambda\times \alpha_{\rm CE}$ during CEE, the initial mass function, the initial separation and
mass, and so on.Considering that BeWDs and their progenitors
locate at the MCs or the Milk Way, we take different metallicities ($Z$=0.0001, 0.004, 0.008 and 0.02) for different galaxies.
In addition, based on the crucial importance  of CEE for binary evolution,
the  effects of combining parameter $\lambda\times \alpha_{\rm CE}$ on BeWD population are discussed.

\subsection{Decretion disk and mass-loss rate of Be star}
Because the decretion disk around Be star is quite complicated, its formation and dynamical evolution still is poorly known \citep{Haubois2012,Rivinius2013}.
Based on a large number of numerical simulations, the decretion disk is fed by material ejected from the Be star,
and it diffuses outwards \citep[e. g.,][]{Rimulo2018,Ghoreyshi2018}. The observational evidences and theoretical estimates
show that the typical mass of a decretion disk is between $10^{-8}$ and $10^{-10}$ M$_\odot$ \citep{Vieira2017,Rimulo2018}.
Based on the theoretical simulations of \cite{Panoglou2016, Ghoreyshi2018},
the mass ejection rate from a Be star to its decretion disk is typically about $10^{-7}$ M$_\odot$ yr$^{-1}$, while
most of ejected materials lose their angular momentum because of themselves interaction, and are re-accreted by the Be star.
The typical mass-loss rate is about $10^{-9}$ M$_\odot$ yr$^{-1}$ \citep{Panoglou2016, Ghoreyshi2018}.

The mass-loss rate of B stars has been theoretically developed by \cite{Vink2000} and \cite{Vink2001},
and can be calculated by stellar parameters (luminosity, effective temperature, metallicity).
Considering the stellar rotation, \cite{Langer1998} gave the mass-loss rate enhanced by
\begin{equation}
\dot{M}_{\rm \Omega}=\left( \frac{1}{1-\Omega/\Omega_{\rm cr}}\right)^\beta\dot{M}_{\rm 0},
\end{equation}
where $\dot{M}_{\rm 0}$ is the mass-loss rate calculated by \cite{Vink2001}, $\beta=0.43$\citep{Langer1998},
$\Omega$ and $\Omega_{\rm cr}$ are the angular velocity and the critical angular velocity, respectively.

The out flow of Be star consists of a slow equatorial decretion disk and a polar wind \citep{Bogovalov2021}.
The mass density of decretion disk can be higher than that of polar wind by two orders of magnitude \citep{Bjorkman1993}.
Following \cite{Lu2009}, we use aspherical structure to simulate the stellar wind of Be star,
and introduce a parameter, $f_{\rm W}$, to describe the mass-loss rate via decretion disk:
\begin{equation}
\dot{M}_{\rm disk}=f_{\rm W}\times\dot{M}_{\rm \Omega}.
\end{equation}
Here, if $f_{\rm W}$=0, there is no decretion disk but a spherical stellar wind; if $f_{\rm W}$=0.9,
there is a decretion disk where 90\% mass of stellar wind is in and the left stellar wind is spherical.
Considering the density structure of stellar wind from Be star \citep{Bjorkman1993}, we take $f_{\rm W}$=0.9 in this work.

For the spherical stellar wind, we use a \cite{Bondi1944} formula to estimate the mass-accretion rate(See details in \cite{Hurley2002}).
For the decretion disk, there is not an accepted accretion model because it depends on the disk structure, dynamical model and
the inclination between the orbital plane and the disk.
It is well known that Be star is misaligned with the orbital plane of BeXB. The main reason is that
the binary system experiences an aspherical symmetry supernova when NS is born.
However, considering that the progenitors of BeWDs do not undergo violent event like supernova,
we assume that the decretion disk is aligned with the orbital plane of BeWD.
Therefore, for simplicity, we assume that WD in the BeWDs can accrete all materials in the accretion disk.
Of course, we can change  $f_{\rm W}$ value to control the ratio of accreted materials to the matter lost by Be star.

\subsection{Method of population synthesis}
With the help of the method of population synthesis which has been used by a series of papers \citep{Lu2009,Lu2012,Lu2013, Zhu2015,Zhu2019,Zhu2021,Han2020},
we simulate the formation and evolution of BeWD's population.
The method of population synthesis involves several input parameters:
the initial mass function (IMF) of the primaries, the mass-ratio distribution of the binaries and the distribution
of initial orbital separations.
Following the previous papers,
we use the IMF of \cite{Miller1979}, a constant mass-ratio distribution,
and assume a flat distribution over $1<\log a_{\rm i}/R_\odot <6$, where $a_{\rm i}$
is the initial orbital separation.
We take $10^8$ initial binary systems with initially circular orbits for each simulation.
Many input parameters (metallicity, alpha value for the CE, accretion efficiency during mass transfer,
the criteria of dynamical mass transfer, et al.) can result in the uncertainties of binary population synthesis (BPS) \citep{Han2020}.
For simplicity, we only consider the effects of the metallicity and the CEE on BeWD population in the present paper.

\section{RESULTS}

As mentioned in the Introduction, the present paper focuses on the
formation channel of BeWDs (Especially, the channel from BesdOB), and the destiny of BeWDs in which WDs accrete
the materials via Be disk. We calculate $10^8$ binary evolution with different metallicity ($Z=0.0001, 0.004, 0.008$ and 0.02) and
combining parameters ($\lambda\times \alpha_{\rm CE}=0.25, 0.5$ and 1.0).
The effects of $\lambda\times \alpha_{\rm CE}$ is only discussed in the next subsection. If it is not specially
mentioned, $\lambda\times \alpha_{\rm CE}$ is always taken as 0.5.

\subsection{Progenitors of BeWDs}
\label{sec:prog}
\begin{figure*}
\begin{tabular}{cc}
\includegraphics[totalheight=3.2in,width=3.5in,angle=-90]{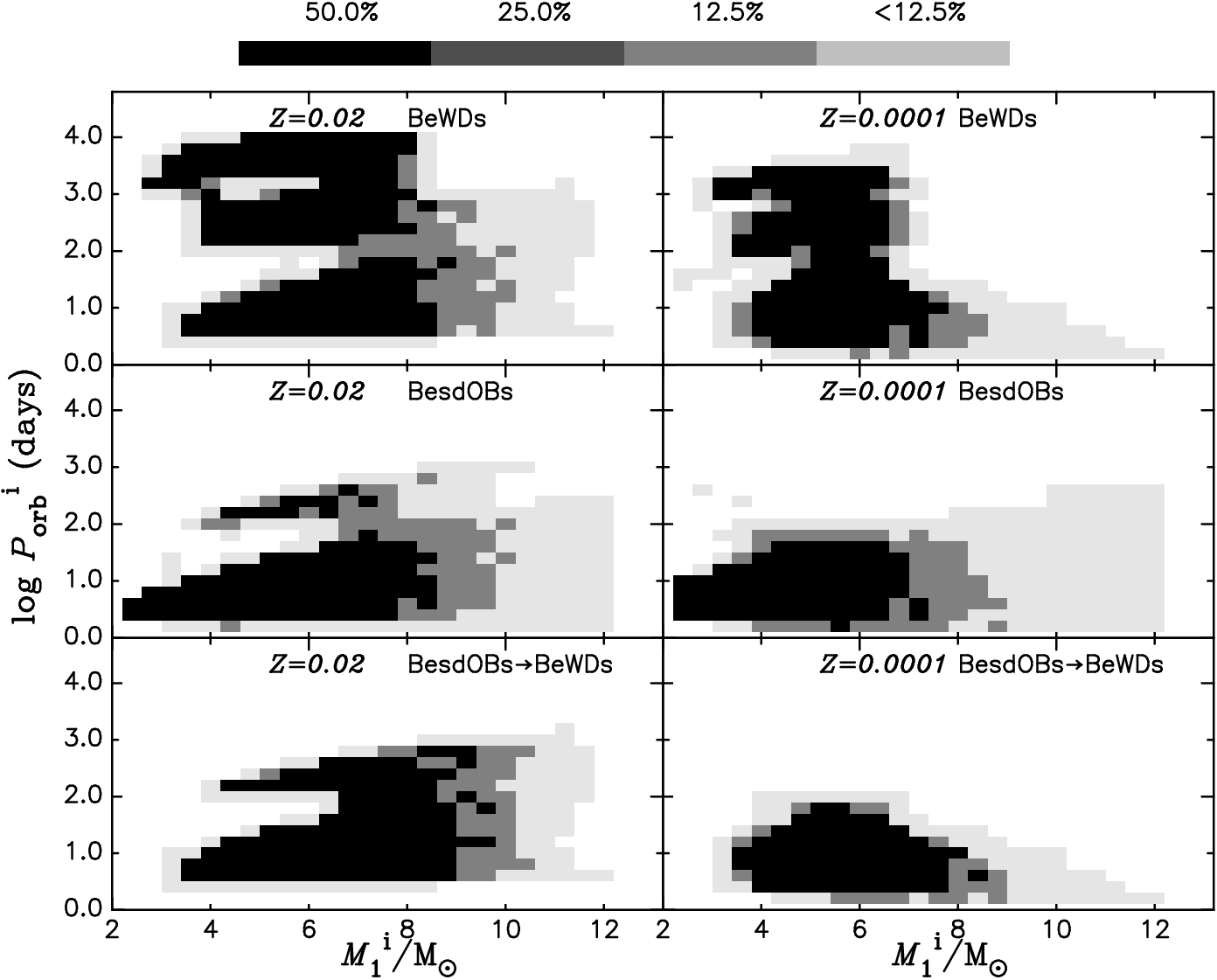}&
\includegraphics[totalheight=3.2in,width=3.5in,angle=-90]{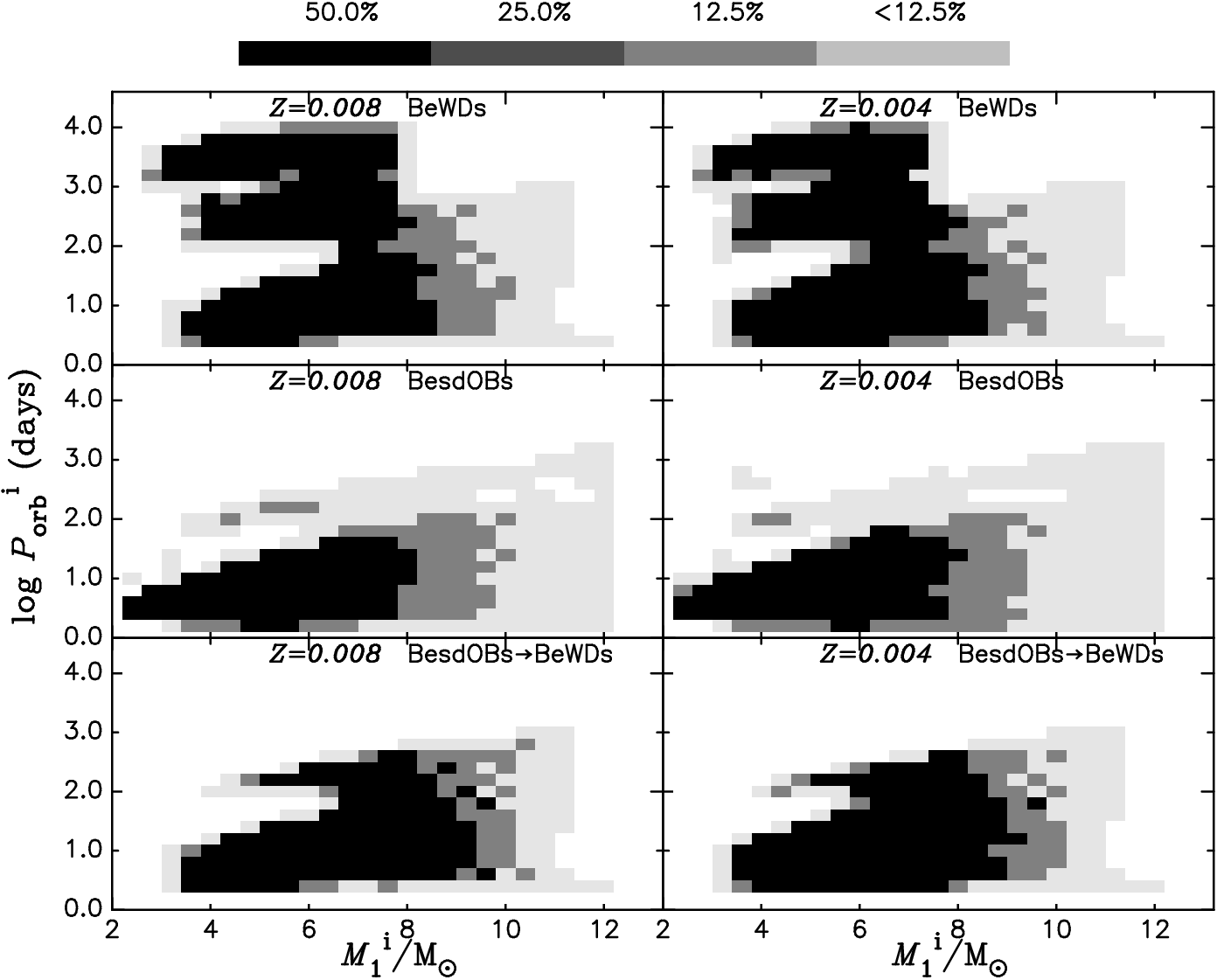}\\
\end{tabular}
\caption{The initial primary masses vs. the initial orbital periods for the progenitors of
         BeWDs (top panels), BesdOBs (middle panels) and those binaries in which BesdOBs can evolve into BeWDs (bottom panels).
         Metallicities in different simulations are given in the top-middle zone of each panel.  }
\label{fig:m1poi}
\end{figure*}

In all $10^8$ initial binary systems, about 4.1\% ($Z=0.02$) - 3.7\% ($Z=0.0001$) of them can evolve into BeWDs, and
about 4.4\%($Z=0.02$) - 6.2\%($Z=0.0001$) of them can become BesdOBs. About 33\% ($Z=0.0001$) - 51\% ($Z=0.002$) of BesdOBs can evolve
into BeWDs. It means that BesdOBs are the most important progenitors of BeWDs.
The average lifetime of BeWDs in our simulations is between about 25 ($Z=0.02$) and 20 ($Z=0.0001$) Myr.
Following \cite{Shao2014},  a constant star formation rate of 5 M$_\odot$yr$^{-1}$ in the Galaxy is taken.
We can estimate that there are about $4.0\times10^5$ BeWDs in the Galaxy.
This estimated number is approximately consistent with
the results in \cite{Shao2014}.
The above uncertainty originates from metallicity. Low metallicity results in a low opacity.
Therefore, the stars with low metallicity have radius smaller than those with high metallicity.
Long orbital period is unfavor for the formation of BeWDs and BesdOBs, as shown by
Figure \ref{fig:m1poi} which gives the distributions of the initial primary masses and the initial orbital periods
for the progenitors of BeWDs, BesdOBs and those binaries in which BesdOBs can evolve into BeWDs.

The initial masses of primaries in the progenitors of BeWDs (the initial masses of WD's progenitors)
are between about 2 and 12 M$_\odot$, and mainly are located between about 3 and 8  M$_\odot$.
The main reason is that these primaries must transfer enough mass so that their companions can be spun up into Be stars whose
masses are higher than 3 M$_\odot$, and themselves can  evolve into WDs.
The initial orbital periods of BeWDs' progenitors mainly distribute three zones
noted by two gaps around $\sim P_{\rm orb}^{\rm i}=10^3$  and $10^2$ days.
The binary systems in which $P_{\rm orb}^{\rm i}>\sim 10^3$ days undergo CEE
when the primaries evolve into AGB, those in which $10^2<P_{\rm orb}^{\rm i}<\sim 10^3$ days do it
when the primaries evolve into Hertzsprung gap or FGB, while those with  $P_{\rm orb}^{\rm i}<\sim 10^2$ days
undergo stable mass transfer when the primaries fill their Roche lobes during the MS phase.
These progenitors with $P_{\rm orb}^{\rm i}<\sim 10^2$ days  first turn into BesdOBs, and then evolve into BeWDs.
Another ranges from hundreds to thousands of day.  The detailed evolution appears in the third subsection.

\begin{figure}
\includegraphics[totalheight=4.in,width=3.5in,angle=-90]{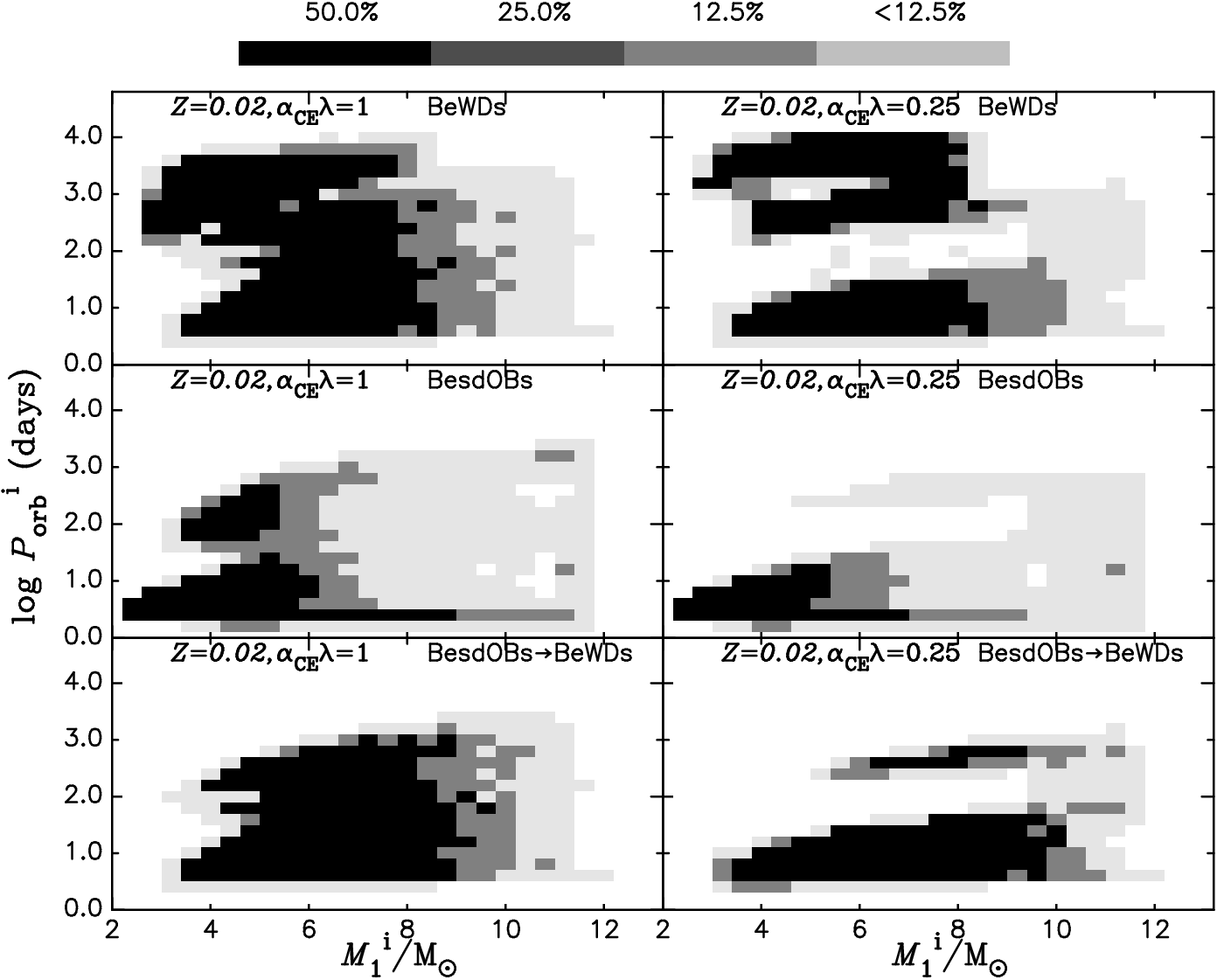}
\caption{Similar to Figure \ref{fig:m1poi}, but for the different $\alpha_{\rm CE}\times\lambda$ during CEE.}
\label{fig:cempi}
\end{figure}

Besides the metallicity, CEE also has great effects on the BeWDs. In this work, we calculate the
effects of combining parameter $\lambda\times \alpha_{\rm CE}$ on BeWD and BesdOB binaries.
The $\lambda\times \alpha_{\rm CE}$ is taken as 0.25, 0.5 and 1.0 in the models with $Z=0.02$, respectively.
We find that the fraction of the binary systems evolving into BeWD are 3.5\%, 4.1\% and 4.6\%.
The larger  $\lambda\times \alpha_{\rm CE}$ is, the more easily the CE is ejected, that is,
binary systems just forming WD can more easily survive after CEE.
Combining Figures \ref{fig:m1poi} and \ref{fig:cempi},
CEE mainly affects the evolution of binary systems in which $10^2<P_{\rm orb}^{\rm i}<\sim 10^3$  days.
As the mentioned in the last paragraph, these binary systems can undergo CEE. When $\lambda\times \alpha_{\rm CE}$ is small,
the binary systems hardly survive because the proportion of orbital energy used to eject CE is too small.
Therefore, these binary systems can not evolve into BeWDs when $\lambda\times \alpha_{\rm CE}=0.25$.

\begin{figure*}
\begin{tabular}{cc}
\includegraphics[totalheight=3.2in,width=3.5in,angle=-90]{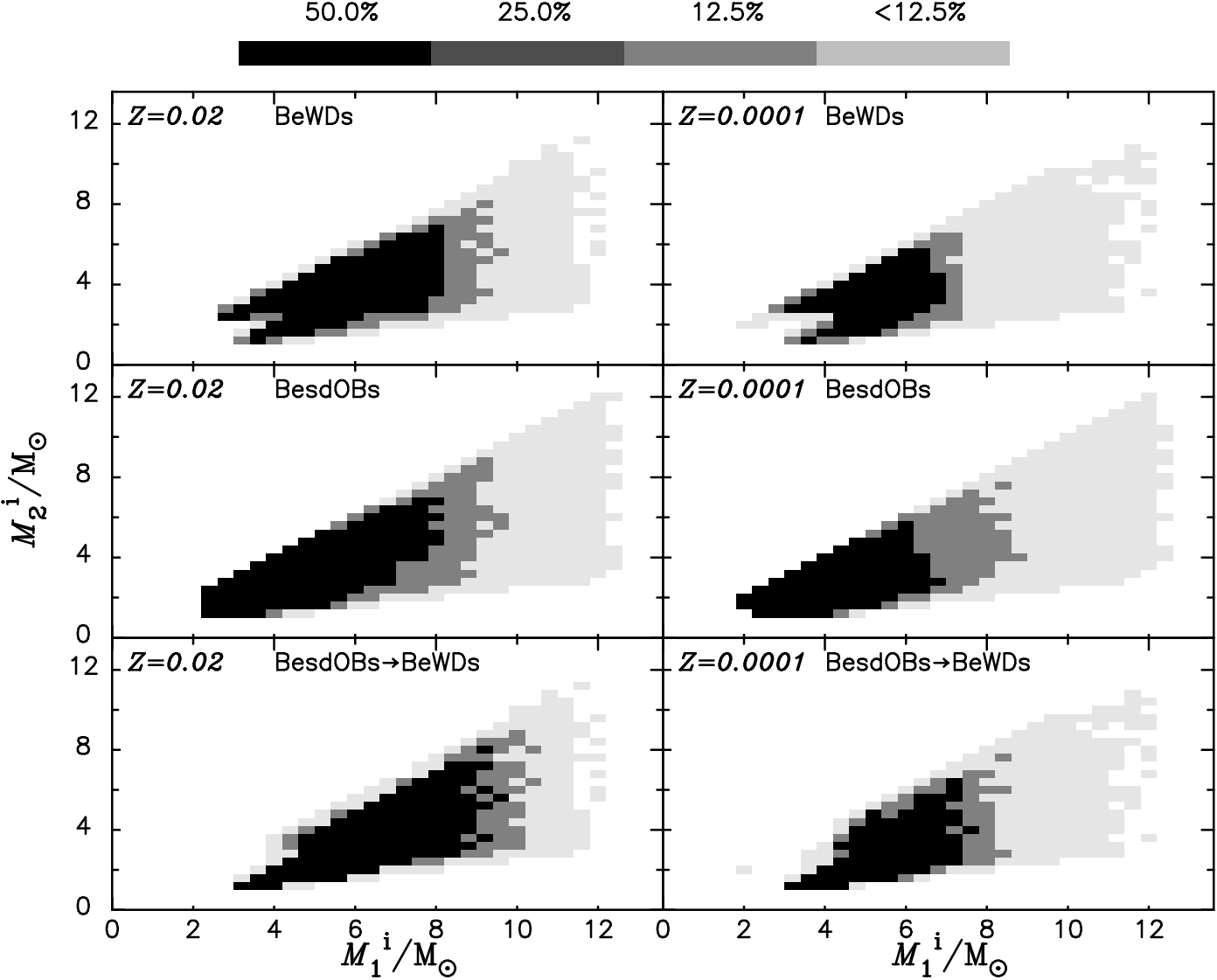}&
\includegraphics[totalheight=3.2in,width=3.5in,angle=-90]{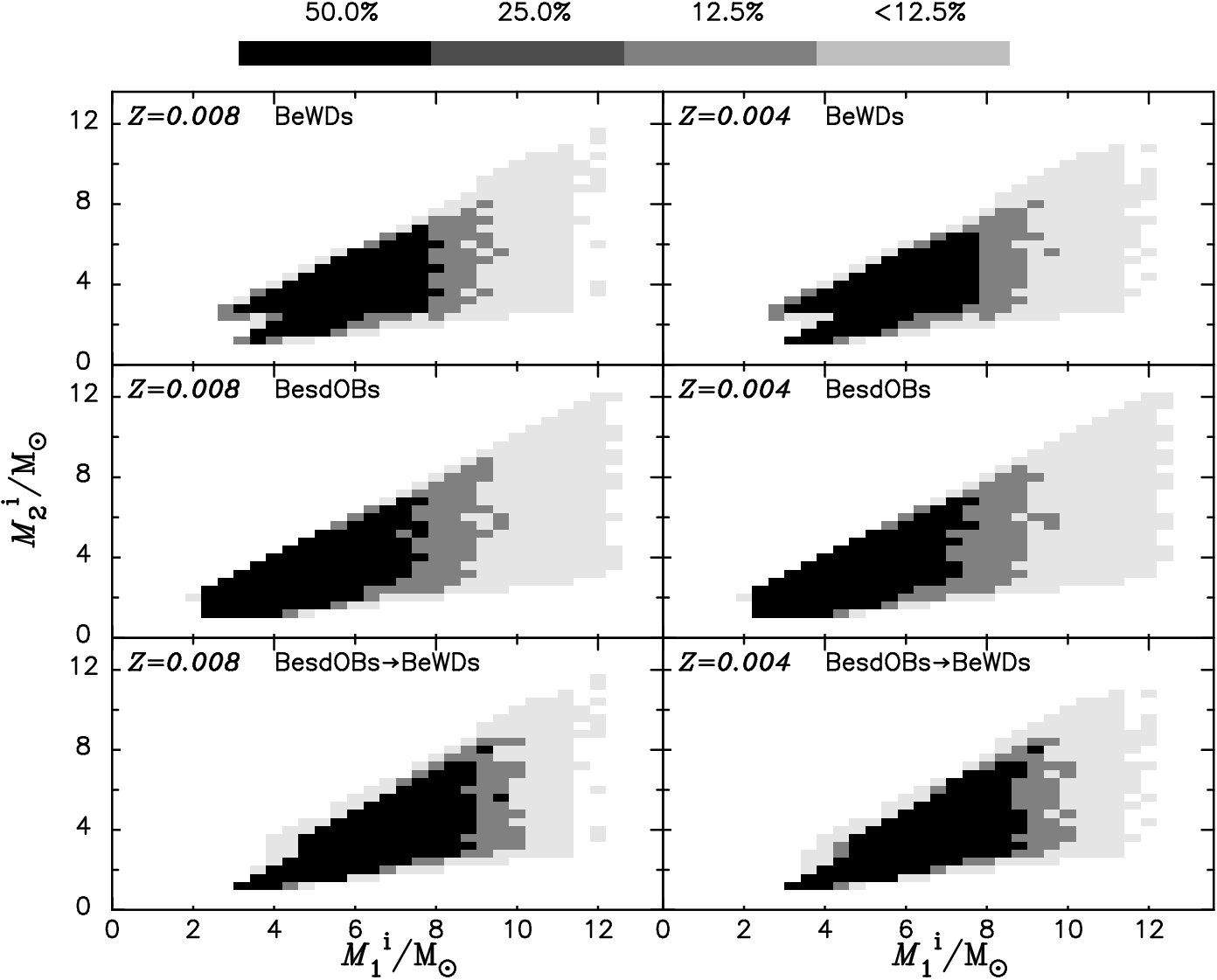}\\
\end{tabular}
\caption{Similar to Figure \ref{fig:m1poi}, but for the initial primary masses vs. secondary masses.
       }
\label{fig:m1m2i}
\end{figure*}

Compared with the BeWD progenitors, those of BesdOBs have shallower distribution of the initial orbital periods because
the primaries must lose their H-rich envelope before the He in the core is exhausted;
they have wider distribution of the initial primary masses because sdOBs can evolve into NSs or even black holes.
As Figure \ref{fig:m1m2i} shown, the initial masses of secondaries have a similar distribution.

\begin{figure*}
\begin{tabular}{cc}
\includegraphics[totalheight=3.2in,width=3.5in,angle=-90]{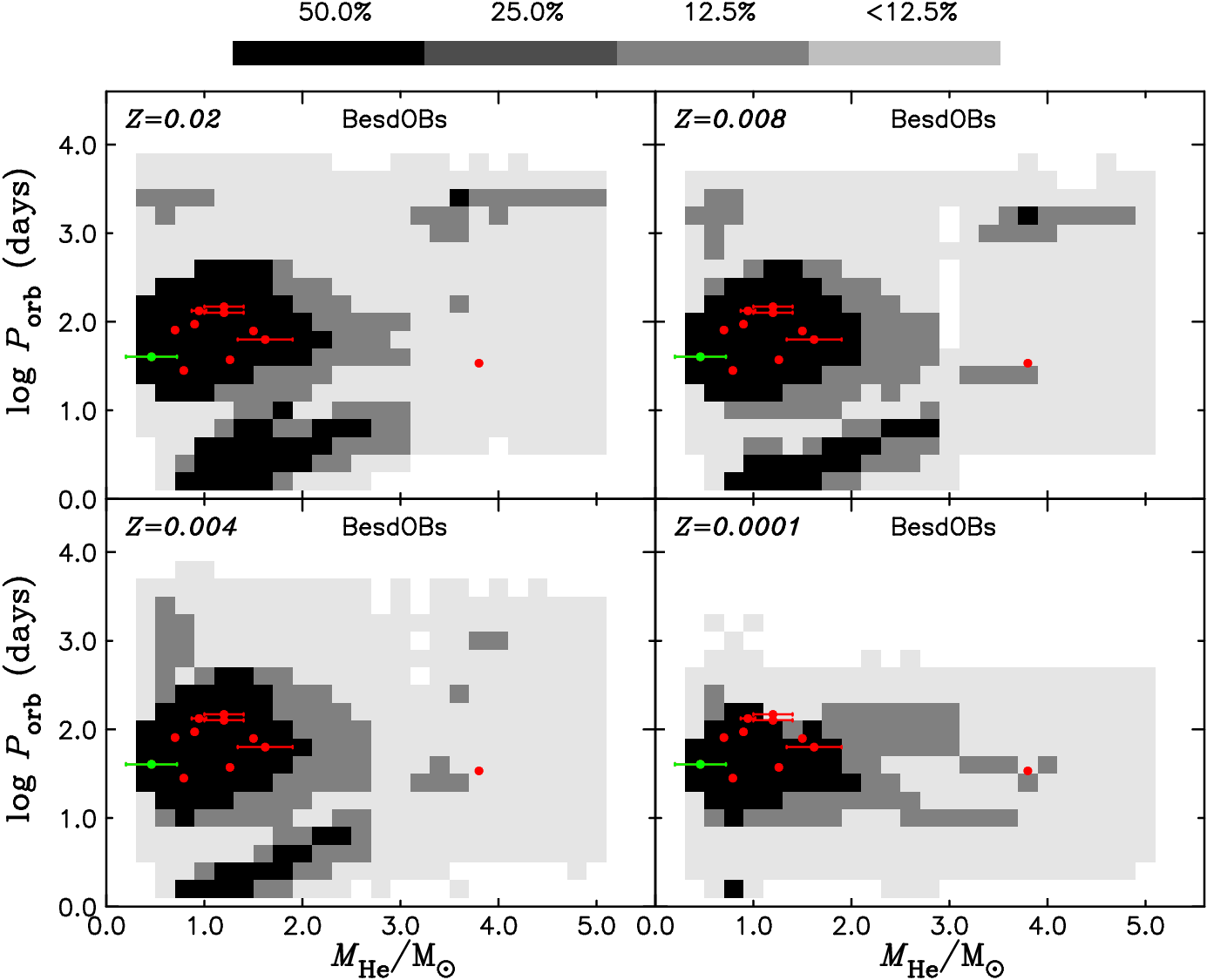}&
\includegraphics[totalheight=3.2in,width=3.5in,angle=-90]{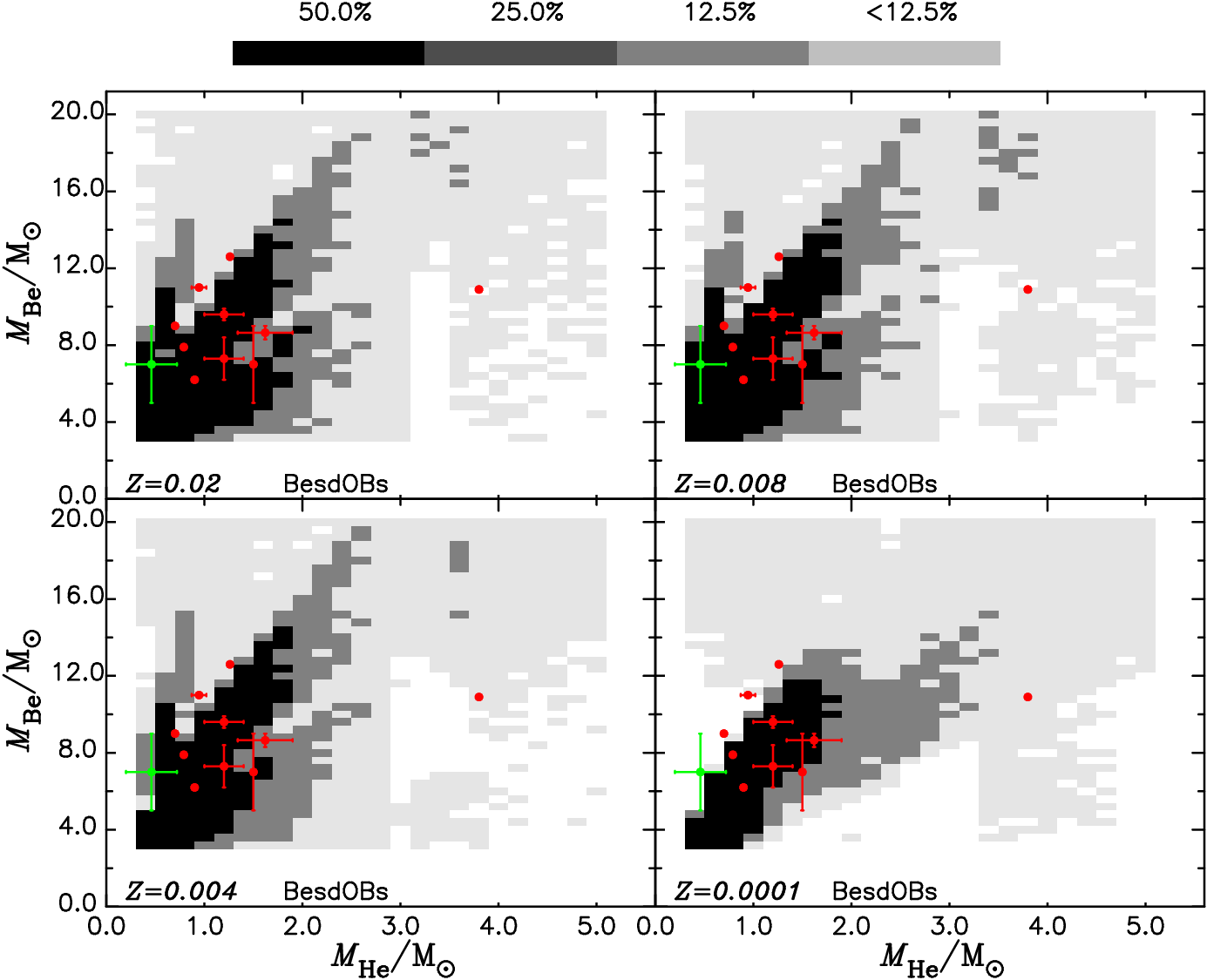}\\
\end{tabular}
\caption{The sdOB masses vs. the orbital periods (left sub-figure) and the sdOB masses vs. Be stars masses (right sub-figure) in BesdOBs.
         The observations listed in Table \ref{tab:Besd} are showed by red points, and HR 6819 is given by green point (See text).}
\label{fig:hem1}
\end{figure*}

As the important progenitors of BeWDs, BesdOBs also are very interesting.
Figure \ref{fig:hem1} shows the distributions of sdOB masses and Be star masses with the orbital periods in
BesdOBs. In our simulations, there are two peaks for the distribution of orbital periods. One is at about
3 days, and these BeWDs have undergone CEE. Another is at about 40 days, and most of them have undergone the
stable mass transfer via Roche lobe.
In the Galaxy, there are about 11 known BesdOBs or candidates whose orbital periods are observed.
As Figure \ref{fig:hem1} shown, these known BesdOBs and candidates are covered by simulated BesdOBs with long orbital periods.
On the observations, there is lack of BesdOBs with the orbital periods between about 2 and 10 days.
One main reason is that the orbital periods are too short so that Be stars soon fill their Roche lobes and BesdOBs undergo the
second CEE. Another is that Be disk hardly form within so short orbital periods.
Theoretically, the He star with a mass higher than about 1.7 M$_\odot$ finally evolves into a NS or a black hole \citep[e. g.,][]{Hurley2000}.
Except $\chi$ Oph, the masses of sdOBs in 11 known BesdOBs are lower than about 1.7 M$_\odot$,
and they have orbital periods longer than about 28 days.
Therefore,  these BesdOBs will become BeWDs.
In our simulations, about 25\% of BesdOBs have massive He stars with a mass higher than 1.7 M$_\odot$. They can evolve
into Be/X-ray binaries if the binaries can survive after a supernova.
If the Be star companion in $\chi$ Oph is a sdOB star, and its mass is about 3.8 M$_\odot$ \citep{Harmanec1987},
$\chi$ Oph may evolve into a Be/X-ray binaries.

\begin{figure*}
\begin{tabular}{cc}
\includegraphics[totalheight=3.2in,width=3.5in,angle=-90]{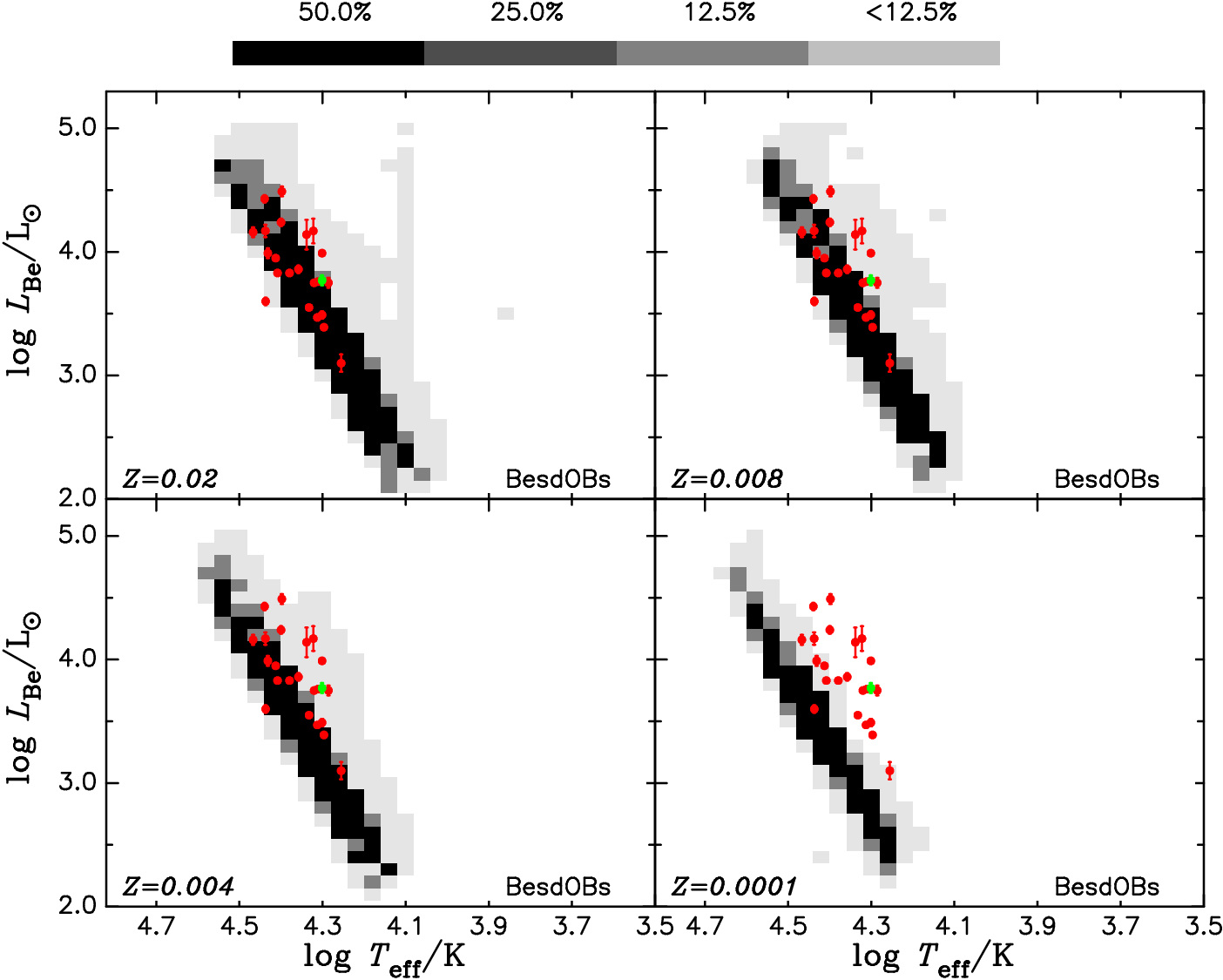}&
\includegraphics[totalheight=3.2in,width=3.5in,angle=-90]{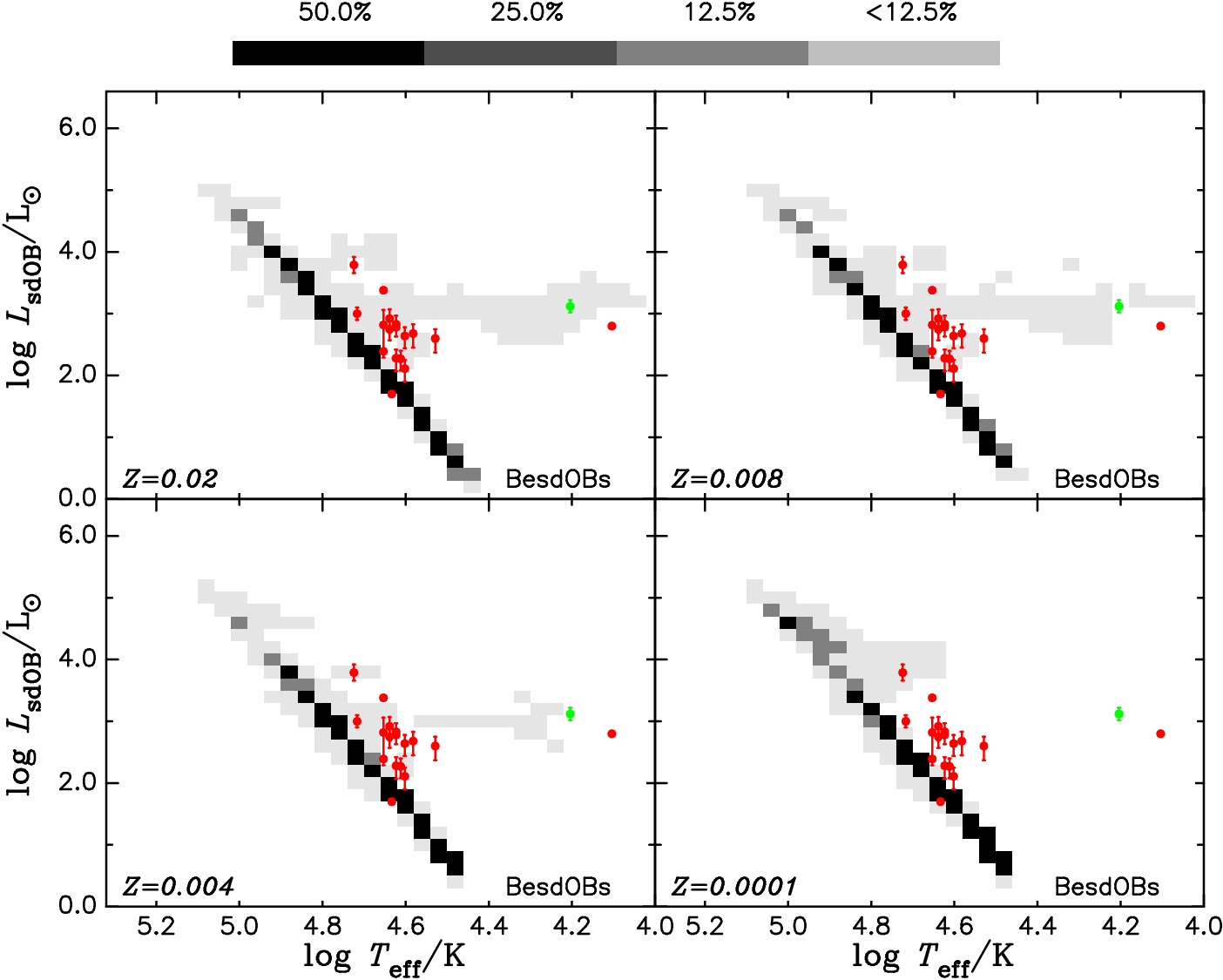}\\
\end{tabular}
\caption{Similar to Figure \ref{fig:hem1}, but for HR diagrams of Be stars (left sub-figure) and sdOBs (right sub-figure) in BesdOBs. }
\label{fig:hehr}
\end{figure*}

Figure \ref{fig:hehr} gives HR diagrams of Be stars and sdOBs in BesdOBs.
Be stars in the known BesdOBs are covered very well by our simulations with high metallicity.
However, most of the sdOBs are located at region where the possibility of forming BesdOBs is very low, especially HR 6819 and ALS 8775.
The main reason is that the luminosity and the effective temperature in our models only depend on the mass of He star and evolutionary time\citep{Hurley2000},
which  results in a shallow region in HR diagram.
HR 6819 is a very intriguing object. \cite{Rivinius2020} suggested that HR 6819 is a triple system which
is composed of a close inner binary consisting a B-type giant plus a black hole and an outer Be star with a wide orbit.
However, according to an orbital analysis, \cite{Bodensteiner2020} considered that HR 6819 is a BesdOB, which
is supported by a new high-angular resolution observations in \cite{Frost2022}.
Based on the model of binary evolution, \cite{Bodensteiner2020} suggested that sdOB in HR 6819 just evolves to higher effective
temperature after mass transfer via Roche lobe, and HR 6819 is very rare.
Similarly, ALS 8775, also called as LB-1, was considered as a binary consisting a B-type star in a 79-day orbit with an
about 70 M$_\odot$ black hole \citep{Liu2019}. However, \cite{Shenar2020} suggested that ALS 8775 comprises a Be star and
a stripped star, that is, it is a BesdOB.
Compared with sdOBs in other known BesdOBs, sdOBs in HR 6819 and ALS 8775 have very low effective temperatures.
As mentioned as \cite{Bodensteiner2020}, these sdOBs just lose their hydrogen-rich envelope, and are evolving into higher effective temperature.
Their results are consistent with ours.

\subsection{BeWD population}
\begin{figure*}
\begin{tabular}{cc}
\includegraphics[totalheight=3.2in,width=3.5in,angle=-90]{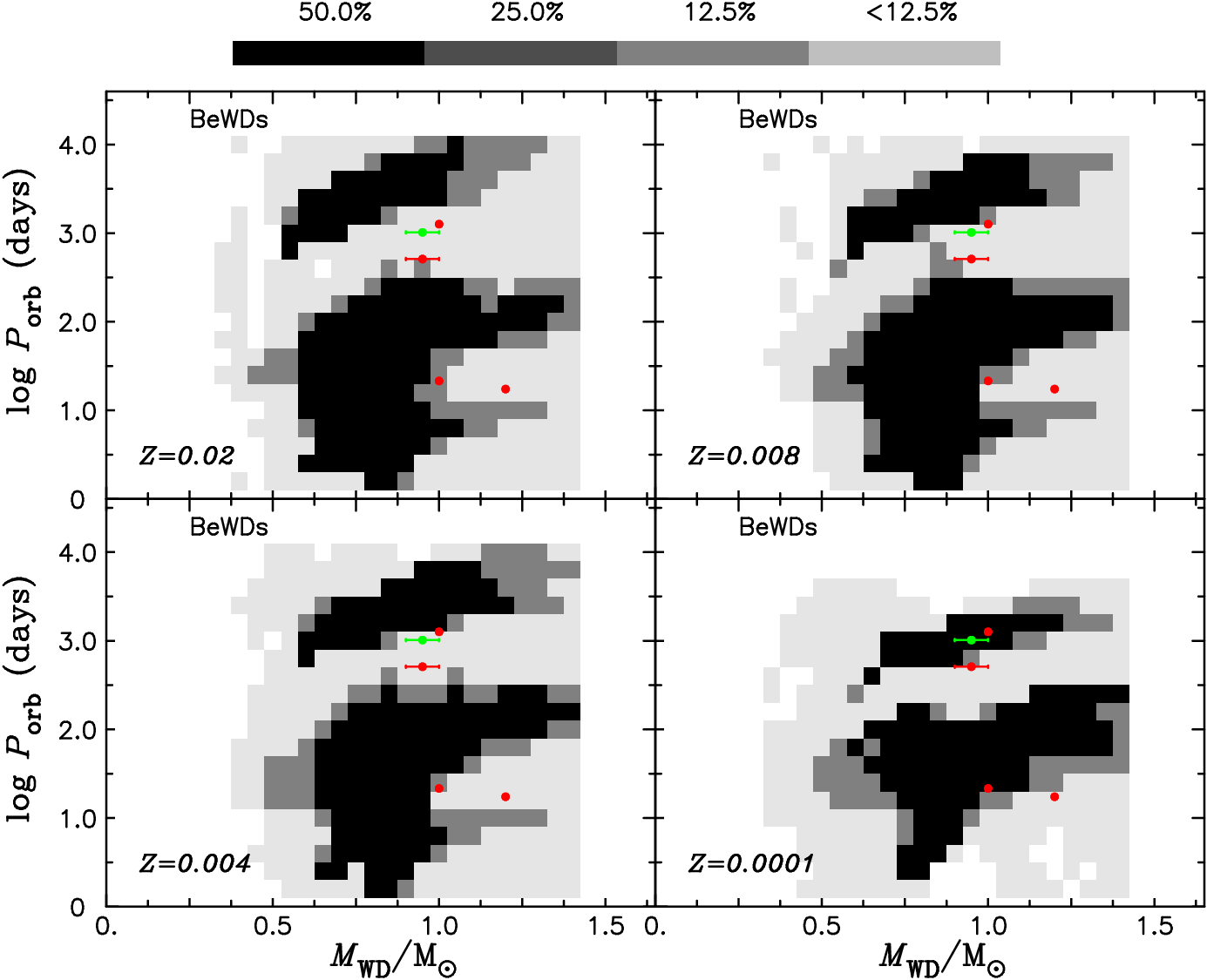}&
\includegraphics[totalheight=3.2in,width=3.5in,angle=-90]{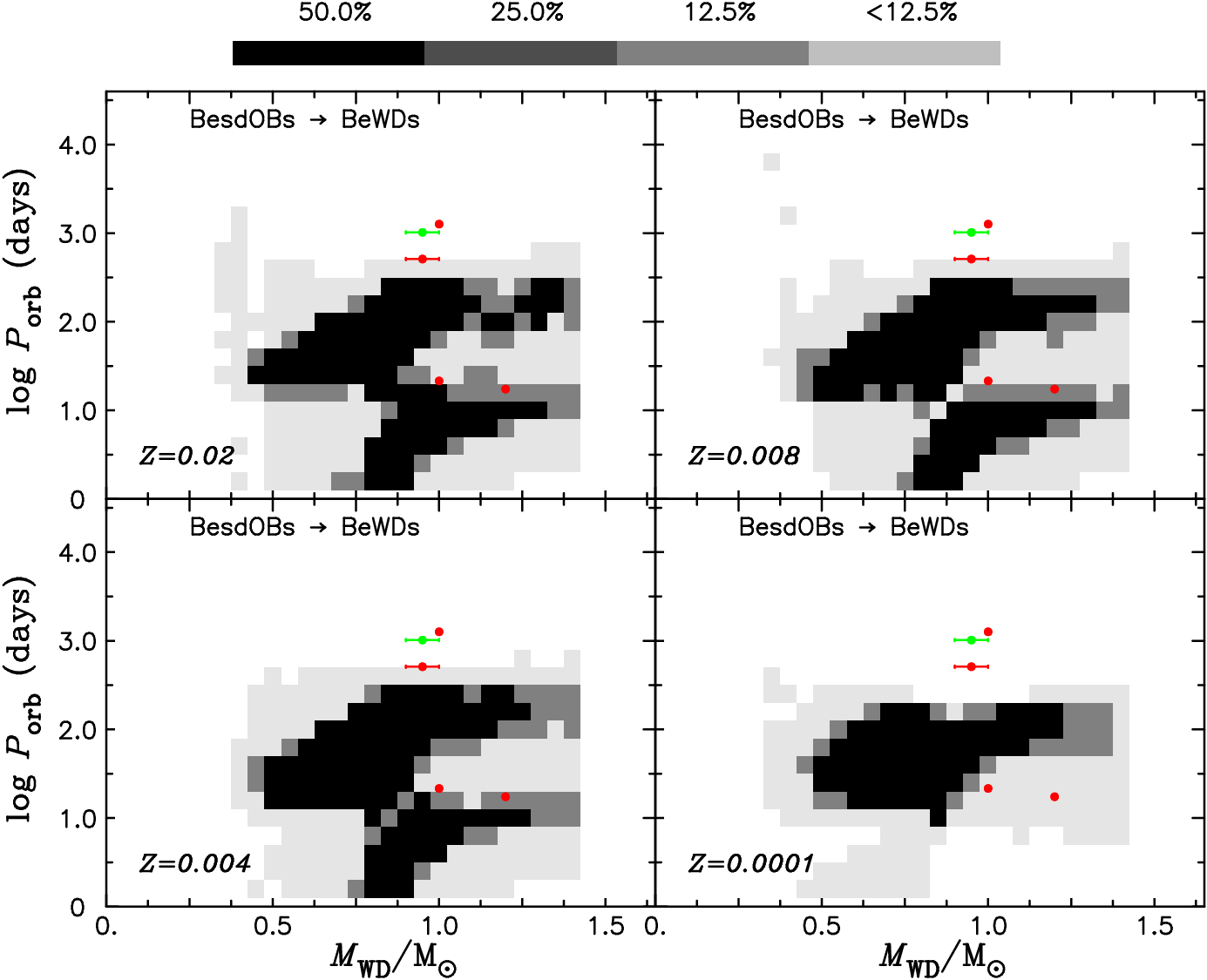}\\
\end{tabular}
\caption{The WD masses vs. the orbital periods for BeWDs (left sub-figure) and the BeWDs which come from BesdOBs (right sub-figure).
         The red points are the known BeWDs which are listed in Table \ref{tab:BeWD}. On observations,
         the orbital period of XMMU J052016.0-692505 is uncertain, and it has two possible values. The longer orbital period (1020 days)
         is given by green point.   }
\label{fig:m1pot}
\end{figure*}

All 6 known BeWDs are detected in the MCs.
Considering that metallicity in a galaxy is not well-distributed,
we plot BeWDs in all simulations with different metallicities.
As illustrated in Figure \ref{fig:m1pot}, our results can cover well 4 BeWDs with known
WD's mass and orbital period.
The BeWDs with orbital periods shorter than about 300 days (such as SWIFT J011511.0-725611 and  J004427.3734801)
have undergone the BesdOB phase, while the left BeWDs (such as XMMU J052016.0-692505 and J010147-715550) experience
a CEE till their primaries evolve into asymptotic giant branch.

\begin{figure*}
\begin{tabular}{cc}
\includegraphics[totalheight=3.2in,width=3.5in,angle=-90]{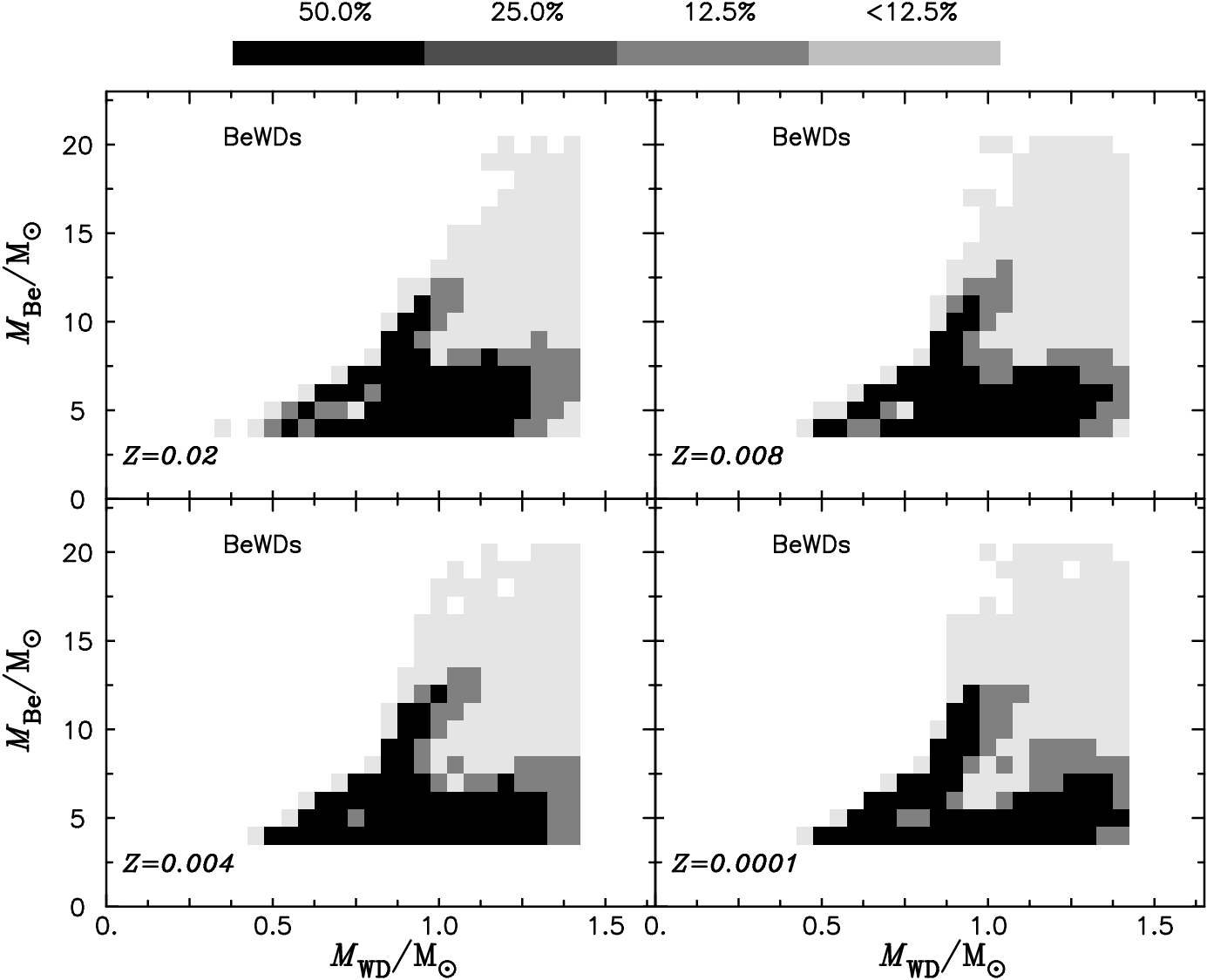}&
\includegraphics[totalheight=3.2in,width=3.5in,angle=-90]{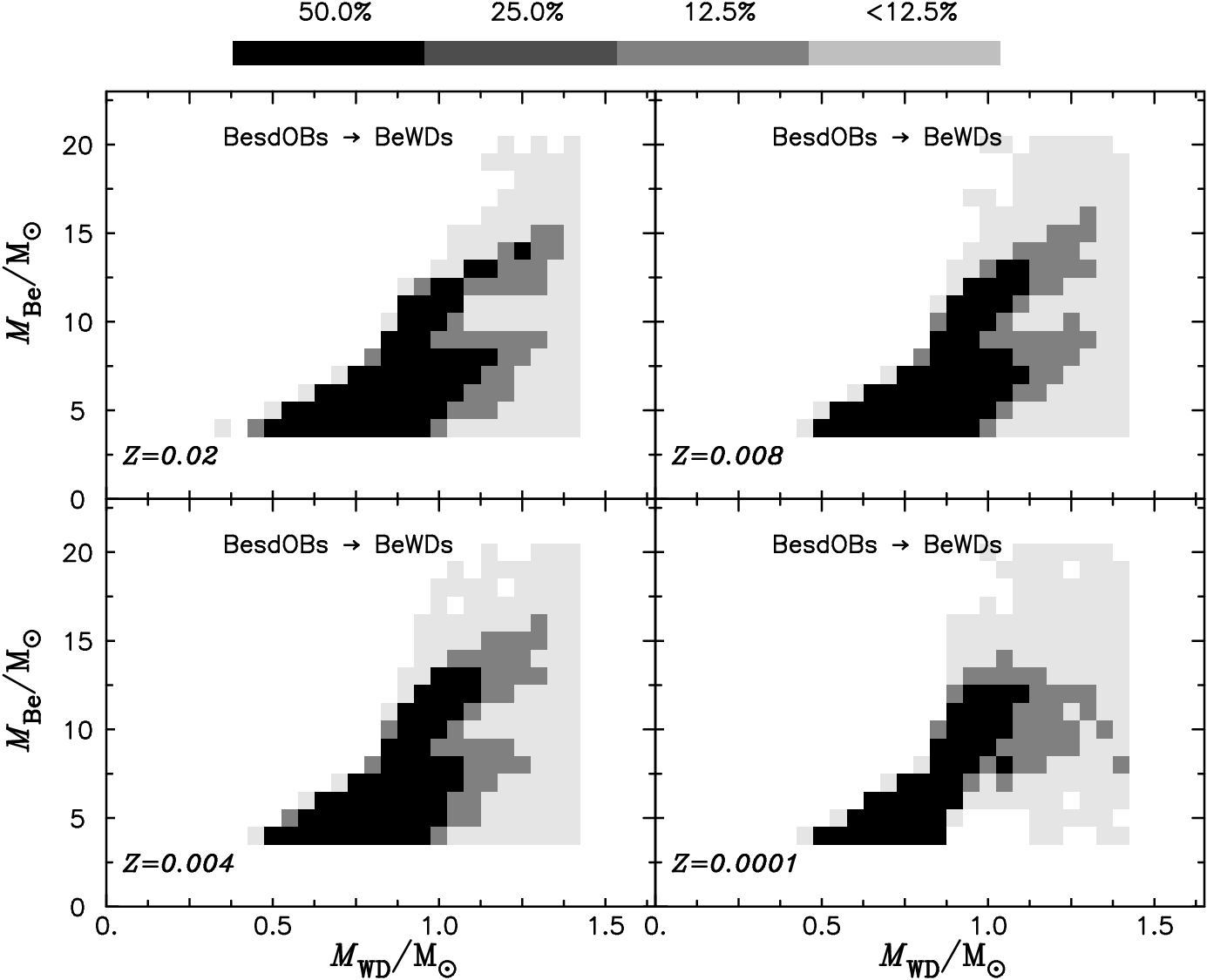}\\
\end{tabular}
\caption{Similar to Figure \ref{fig:m1pot}, but for the masses of WDs Be stara in BeWDs. }
\label{fig:m1m2t}
\end{figure*}

Similar result appears in Figure \ref{fig:m1m2t} which shows the distributions of WDs' and Be stars' masses in BeWDs.
Obviously, the mass distribution has two zones. One zone has  larger $q$ ($q=M_{\rm Be}/M_{\rm WD}$).
These BeWDs have undergone efficient mass transfer, and BesdOBs can be their progenitors.
Another has low $q$. These BeWDs have undergone inefficient mass transfer.

Observationally, it is difficult to find a BeWD. The optical properties of BeWDs
are similar to those of binaries consisting of Be star and neutron star,
and it is hard to distinguish them\citep{Coe2020}.
Compared with neutron star X-ray binaries, BeWDs have soft X-ray range.
Usually, if the mass-accretion rate is higher than the minimum stable burning rate ($M_{\rm cr}$$\sim10^{-7}$ M$_\odot$yr$^{-1}$),
the accreting WD is a persistent soft X-ray source, or else it is a transient X-ray source during nova burst.
For a persistent soft X-ray source or ones during nova burst, the X-ray luminosity can be estimated by the nuclear burning rate of
accreted material \citep[e. g.,][]{Wolf2013,Chen2015}. During the quiet phase,
the X-ray emission mainly is produced by the gravitational potential released by material accreted.
All known BeWDs are transient X-ray sources. Most of time, they are in the quiet phase.
Following \cite{Chen2022}, although soft X-ray emission can be easily absorbed by the interstellar medium,
the X-ray luminosity of BeWD can be estimated by the mass-accreting rate via
\begin{equation}
L_{\rm X}=\frac{GM_{\rm WD}}{R_{\rm WD}}\dot{M},
\label{eq:Lx}
\end{equation}
where $G$ is gravitational constant.

\begin{figure*}
\begin{tabular}{cc}
\includegraphics[totalheight=3.2in,width=3.5in,angle=-90]{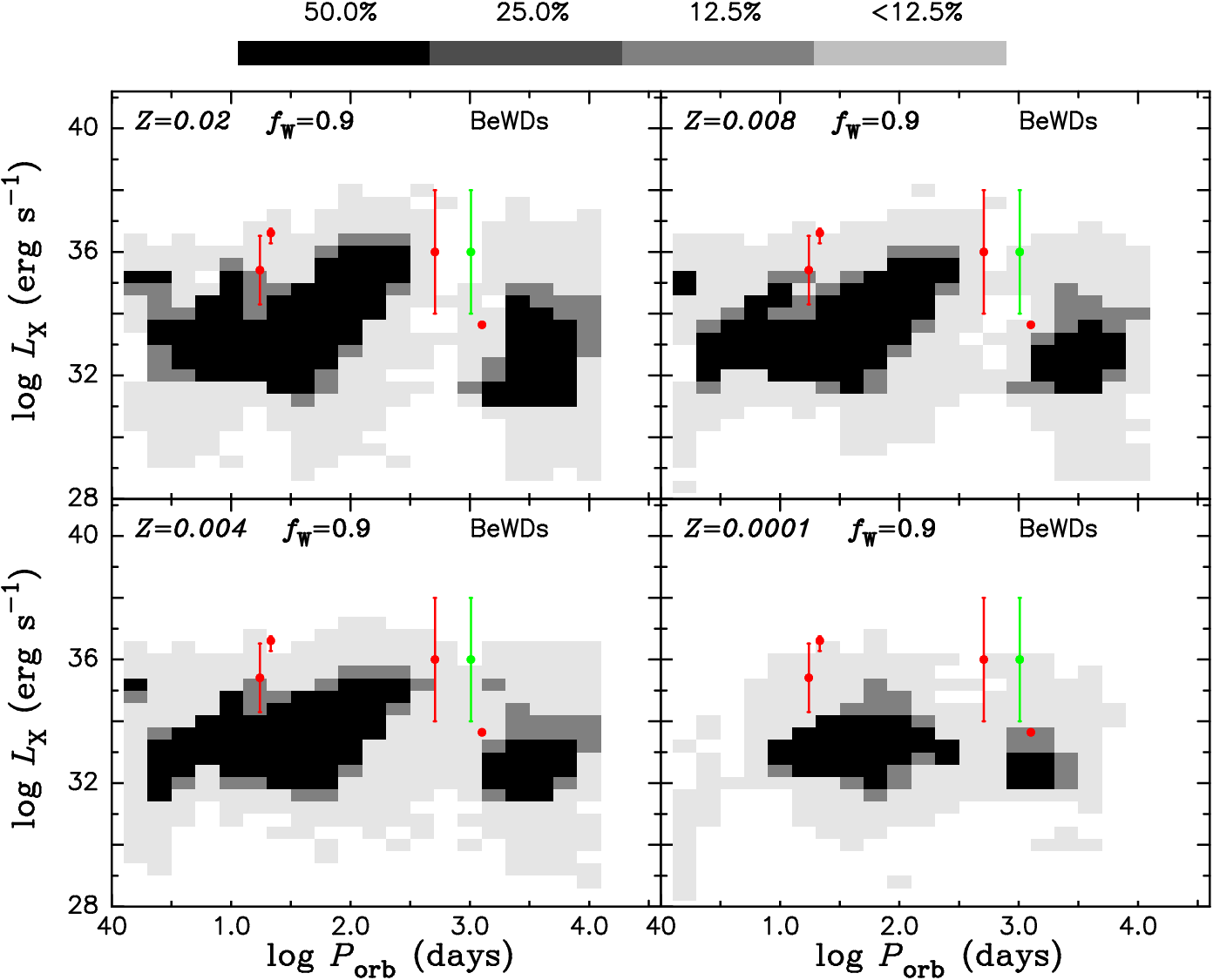}&
\includegraphics[totalheight=3.2in,width=3.5in,angle=-90]{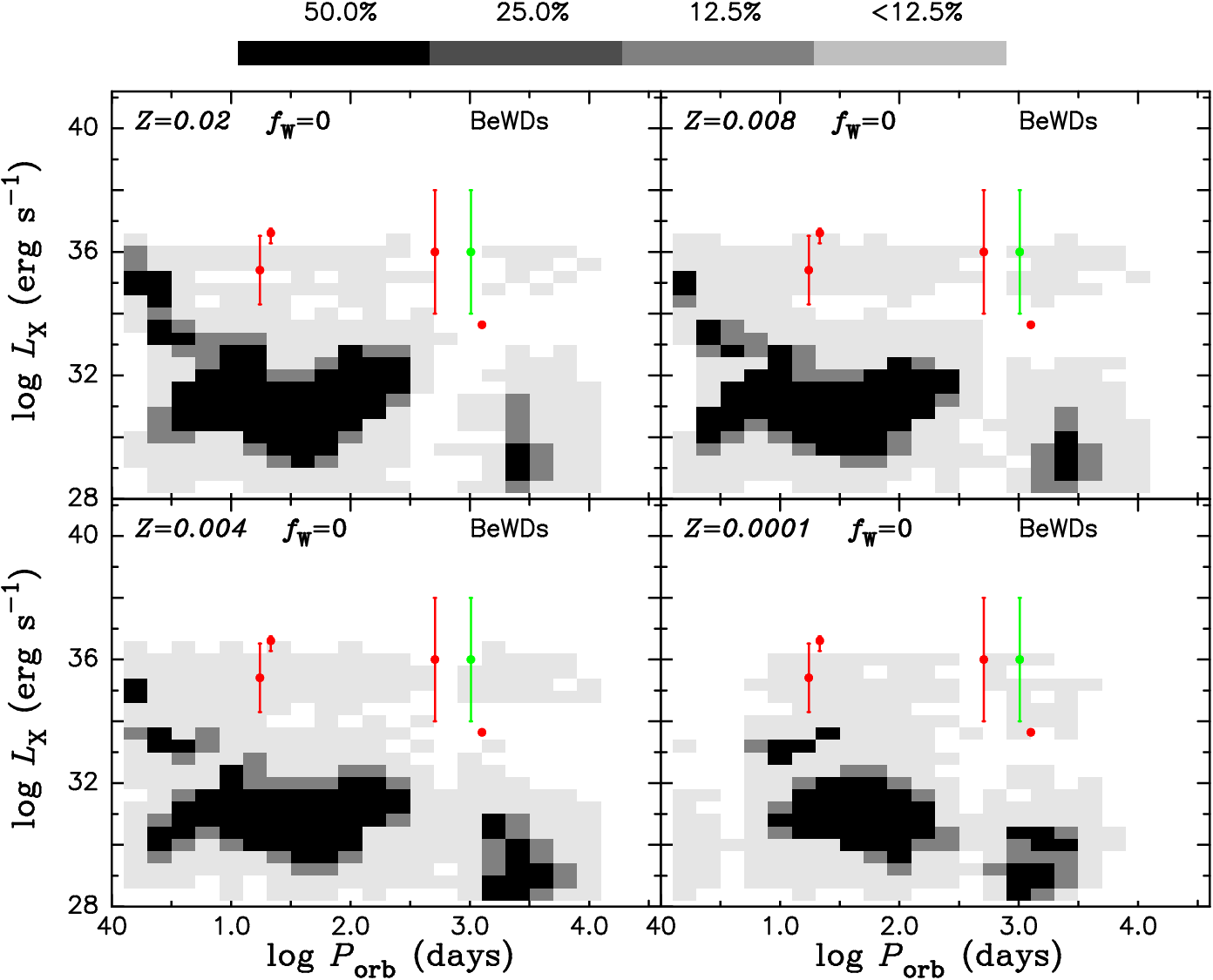}\\
\end{tabular}
\caption{Similar to Figure \ref{fig:m1pot}, but for the orbital periods vs. X-ray luminosities calculated by Eq. (\ref{eq:Lx}).
The left and the right sub-figures represent the simulations of $f_{\rm W}=0.9$ and 0, respectively. }
\label{fig:poma}
\end{figure*}

Figure \ref{fig:poma} gives the estimated X-ray luminosities with different $f_{\rm W}$s.
For the models with $f_{\rm W}$=0 (a spherical stellar wind, the right sub-figure of Figure \ref{fig:poma}),
our results hardly cover the observational samples.
Even for the standard models with $f_{\rm W}$=0.9 (a decretion disk being dominated in the stellar wind, the left sub-figure of Figure \ref{fig:poma}),
the known BeWDs appear in the top region.
There are two possible reasons as follows.
One is that our models underestimate the mass-loss rate of Be stars. \cite{Carciofi2012} suggested
that the mass-loss rate by decretion disk can reach up to about $10^{-7}-10^{-10}$M$_\odot$ yr$^{-1}$ \citep[Also see][]{Rimulo2018,Ghoreyshi2018},
which is higher than times of that observed in B stars \citep{Puls2008}.
Another is that these BeWDs measured X-ray luminosities are on outburst phase. In BeWDs, outbursts are
divided into two types. The type-I outburst originates from the  thermonuclear runaway (TNR).
Usually, the X-ray luminosity produced by TNR in BeWDs is too faint to be detected. The type-II outburst
 is triggered by accretion-disk instability. SWIFT J011511.0-725611 just experienced a type-II outburst \citep{Kennea2021}.
 The X-ray luminosity during type-II outburst can be higher than that during quiet phase
 for at least two orders of magnitude \citep{Kahabka2006,Kennea2021}.
 Therefore, the model with a stellar wind dominated by decretion disk is better.

\subsection{BeWD destiny}
Theoretically, BeWD destiny depends on not only its orbital period and mass ratio ($q=\frac{M_{\rm Be}}{M_{\rm WD}}$), but also  WD types (CO or ONe WD).
In our simulations, the ratios of COWDs to ONeWDs in BeWDs are about 7/3 ($Z=0.02$) and 6/4 ($Z=0.0001$).
This ratio is mainly determined by the initial mass forming CO and ONe WDs. The former range of initial masses in BeWDs is
between about 2 and 6 M$_\odot$, while the later is between about 6 and 12 M$_\odot$.

\begin{figure*}
\begin{tabular}{cc}
\includegraphics[totalheight=3.2in,width=3.5in,angle=-90]{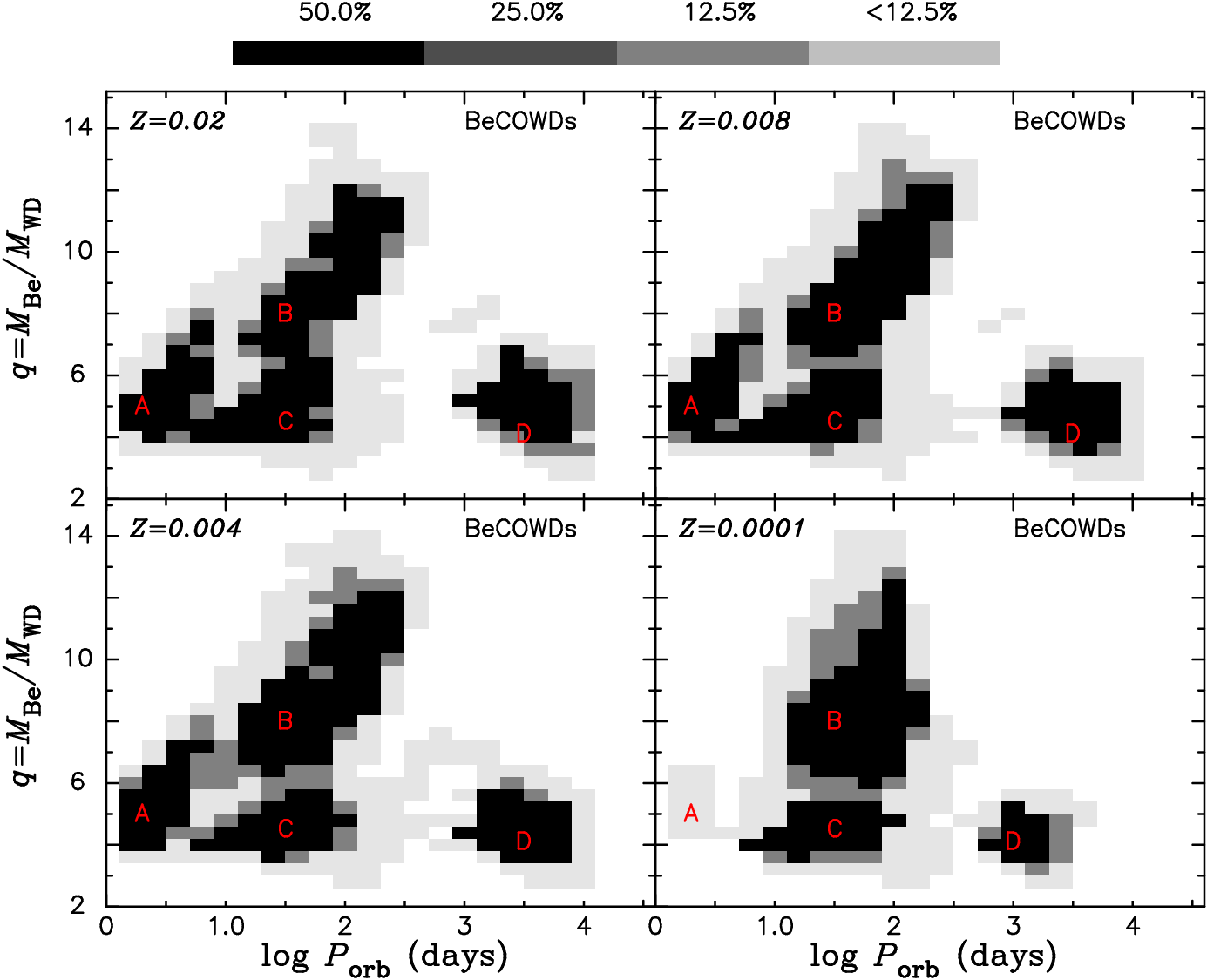}&
\includegraphics[totalheight=3.2in,width=3.5in,angle=-90]{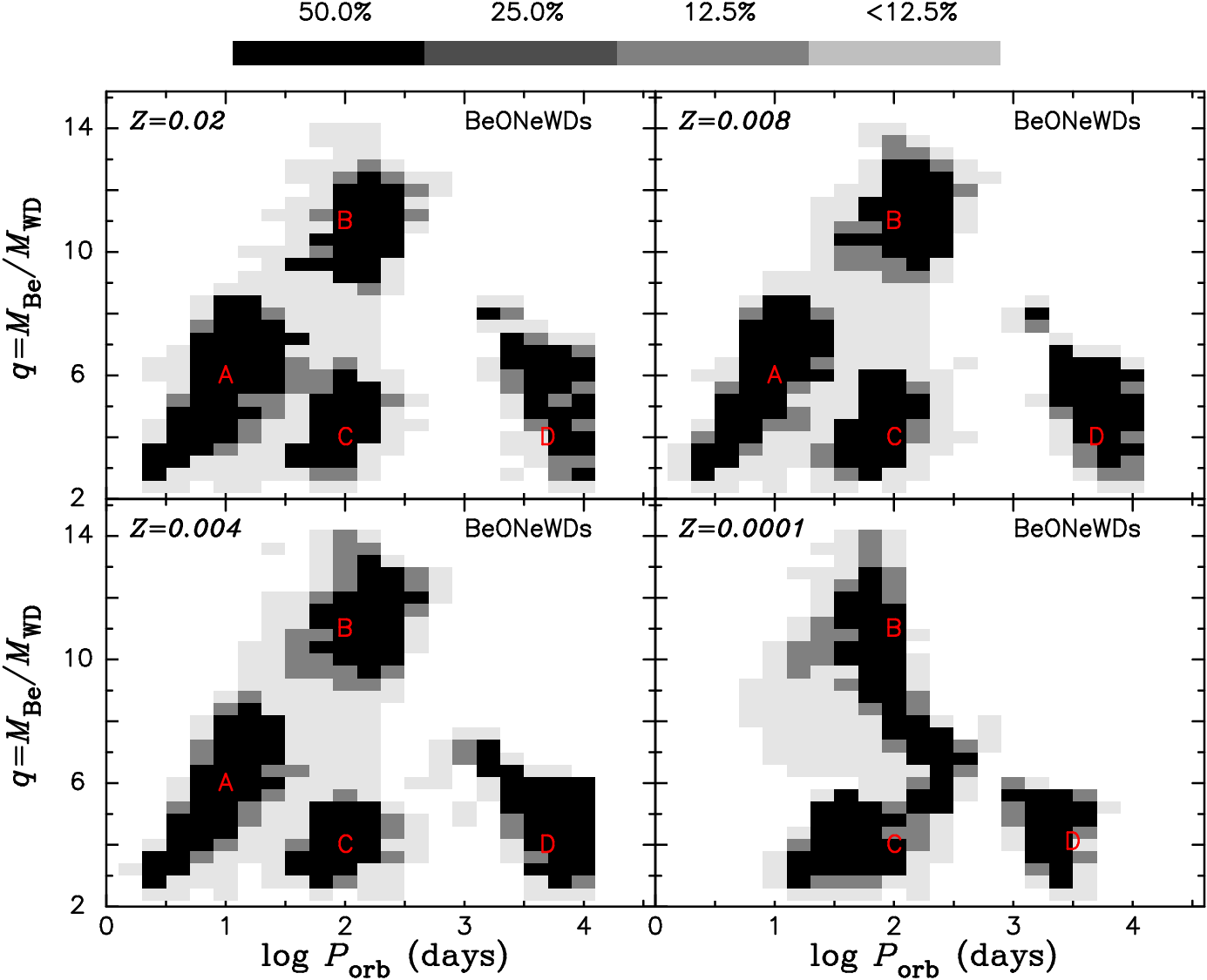}\\
\end{tabular}
\caption{Distributions of the orbital periods vs. the $q (q=\frac{M_{\rm Be}}{M_{\rm WD}})$ for BeCOWDs (left sub-figure) and
         BeONeWDs (right sub-figure). The letters 'A', 'B', 'C' and 'D' represent different evolutionary channels(See text).  }
\label{fig:poq}
\end{figure*}

Figure \ref{fig:poq} shows the distributions of the orbital periods and $q$ for BeCOWDs and BeONeWDs.
There are 4 zones which are noted 'A', 'B', 'C' and 'D'. Every zone represents different evolutionary channels.
The proportions of BeCOWDs produced by A, B, C and D channels are about 20\%, 35\%, 25\% and 20\% in all simulations
except that they are about 5\%, 50\%, 35\% and 10\% in the extremely metal-poor simulation ($Z=0.0001$).
For BeONeWDs, they are about 40\%, 20\%, 20\% and 20\% in the metal-rich models,
while they are about 0, 50\%, 30\% and  25\% in the extremely metal-poor simulation.

As mentioned in subsection \ref{sec:prog},  BeWDs in A and B zones have been BesdOBs.
Differently, the progenitors of BeCOWDs (or BeONeWDs) in A zone have orbital periods whose range is between about 20 (30) and 200 (600) days.
Their primaries fill their Roche lobes on late phase of Hertzsprung gap or even giant phase.
When they evolve into the first giant branch, the progenitors experience CEE.
If the binaries do not merge, the orbital periods shrink to about several percents, the primaries become sdOBs,
and the secondaries are spun up to Be stars via mass accretion before CEE or tidal interaction after CEE.
The progenitors become BesdOBs, and then they evolve into BeWDs with the shortest orbital periods.
The progenitors of BeCOWDs (or BeONeWDs) in B zone have initial orbital periods shorter than 20 (30) days.
Their progenitors fill Roche lobes during main sequence, and undergo a stable and efficient mass transfer so that
the secondary masses exceed primary ones. Even the primaries evolve into the first giant branch,
the mass transfer is still stable, which results in lengthening orbital periods.
For the BeWDs in C and D zones, their progenitors have wide orbits.
The progenitors of BeCOWDs (or BeONeWDs) in C zone have the orbital periods between about 200 (600) and 1000 (2000) days,
their primaries fill Roche lobes till they evolve into early AGB phase. Usually, these binaries undergo CEE.
For the BeWDs in D zone, the orbital periods of their progenitors are so long that the primaries can not fill Roche lobes.
The secondaries are spun via accreting stellar wind.
For very metal-poor models ($Z=0.0001$), A zone lacks BeWDs. The main reason is as follows:  Very low metallicity results in  very
small stellar radius so that the primaries can not fill their Roche lobe.

\subsubsection{Merger of WD and non degenerated star }
Be stars or their successors in BeWDs in the A, B and C zones of Figure \ref{fig:poq}
can fill their Roche lobes with their evolution.
If $q$ is higher than $q_{\rm cr}$, CEE occurs.
It is well known that CEE can produce two different results: merger if the CE can not be ejected, or else a close binary.
The former is usually occurs when Be stars or their successors are on main sequence or Hertzsprung gap,
which is a merger of WD and H-rich star.
The later usually forms the close binary consisting of a WD and a He star.  With He star evolution, it also fills
its Roche lobe. CEE may occur again, which produces a merger of a WD and a He star or double WDs.

\begin{figure}
\includegraphics[totalheight=3.5in,width=3.5in,angle=-90]{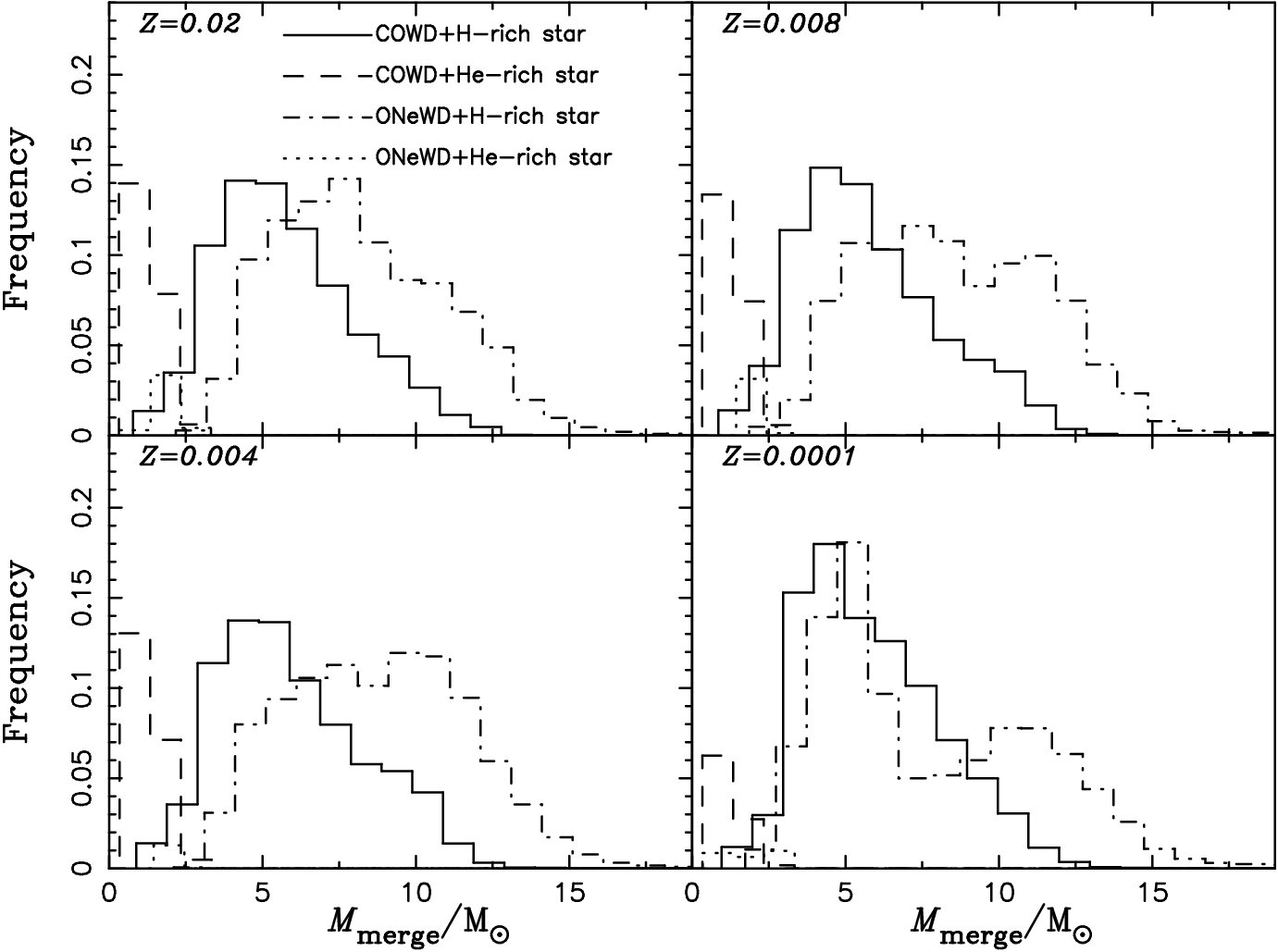}
\caption{Mass distributions of stars produced by merger of a WD and a non degenerated star.
Different lines represent the merger of different types of WD and star, which are given in the top-middle zone.
The frequency of mergers involving COWDs and ONeWDs is normalized to 1, respectively. }
\label{fig:sdm}
\end{figure}

Figure \ref{fig:sdm} shows the mass distributions of stars produced by merger of a WD and a non degenerated star.
Here, we do not consider the mass loss during merger events.
In our simulations, about 60\% ($Z=0.02$) - 70\% ($Z=0.0001$) of BeWDs finally merge into single star systems,
in which about 80\% ($Z=0.02$) - 90\% ($Z=0.0001$) of merger events  involve WD and H-rich star.
These mergers may observationally correspond to luminous red novae\citep{Soker2003,Ivanova2013,Howitt2020}.
The left mergers involving WD and He-rich star may form He giant stars\citep[e g.,][]{Hurley2002}.
Their progenitors have successfully experienced CEE, and the envelope of
Be stars or their successors are ejected. Therefore, the mass of these He giant stars is lower than
about 2.0 M$_\odot$.

\subsubsection{Double white dwarf binaries}
\begin{figure}
\includegraphics[totalheight=3.3in,width=3.5in,angle=-90]{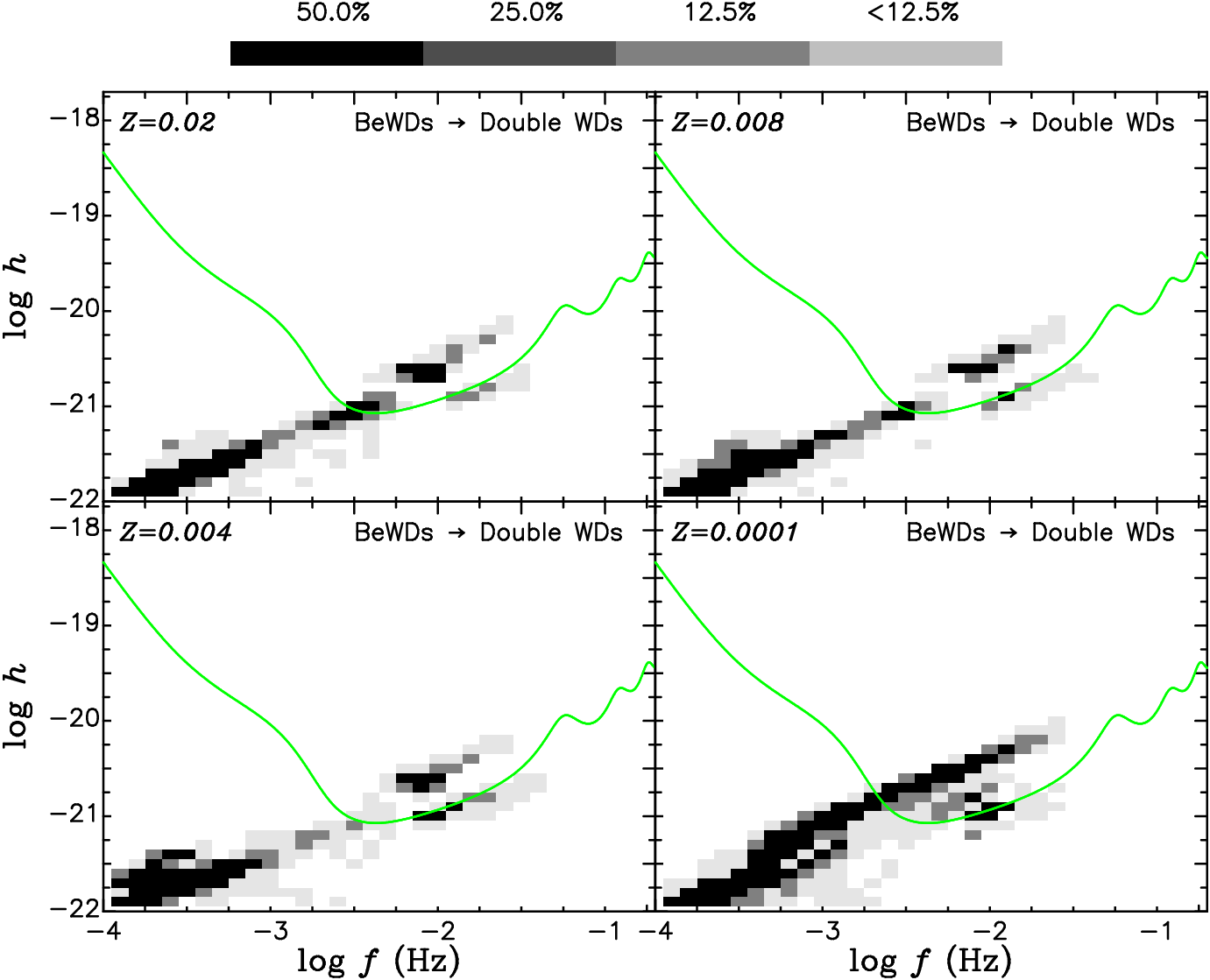}
\caption{Distribution of gravitational sources from double WDs in the strain-frequency space. Green lines
         give the LISA sensitivity.   }
\label{fig:lisa}
\end{figure}

About 30\%-40\% of BeWDs' successors do not merge during CEEs. They usually evolve into double WDs.
They represent the most likely gravitational wave source (GWS) detected
by $LISA$ mission \citep[e .g.][]{Amaro-Seoane2017}.
\cite{Peters1963} gave the GW luminosity of a binary system.
Assuming an orbital period and a sinusoidal wave, the strain amplitude of GW, $h$, can be
given by
\begin{equation}
h=5.0\times10^{-22}\left(\frac{\mathcal{M}}{M_\odot}\right)^{5/4}\left(\frac{P_{\rm orb}}{1 {\rm hour}}\right)^{-2/3}\left(\frac{d}{1 {\rm kpc}}\right)^{-1},
\end{equation}
where $\mathcal{M}=(M_1M_2)^{3/5}/(M_1+M_2)^{1/5}$ is the chirp mass.
Considering that the distances of most double WDs known in the Galaxy is shorter than 1 kpc \citep{Korol2017,Kupfer2018,Burdge2019,Li2020},
we take $d$ as 1 kpc.
Figure \ref{fig:lisa} gives the distribution of GWs from double WDs in the strain-frequency space.
The GWs' frequency detected potentially by LISA are mainly  at about $3\times10^{-3}$ and $\times10^{-2}$ Hz,
which greatly depends on the LISA sensitivity.

\subsection{Supernova Ia progenitors}
Following \cite{Lu2009}, we also structure an aspherical stellar-wind model for BeWDs.
It depends on the mass-accretion rates ($\dot{M}_{\rm a}$) whether an accreting WD's mass can efficiently increase\citep[e. g.,][]{Nomoto2007}.
If $\dot{M}_{\rm a}$ is higher than a critical value ($\dot{M}_{\rm cr}$), the accreted hydrogen materials can steadily burn, or less
TNR can occur on the surface of accreting WD. For the former, the accreted materials can efficiently turn into WD matter after the burning.
For the later, the accreted materials can partly be ejected during the TNR, even WD matter is eroded when $\dot{M}_{\rm a}$ is lower than
about $\frac{1}{8}\dot{M}_{\rm cr}$ \citep{Yaron2005,Lu2006}

\begin{figure*}
\begin{tabular}{cc}
\includegraphics[totalheight=3.2in,width=3.5in,angle=-90]{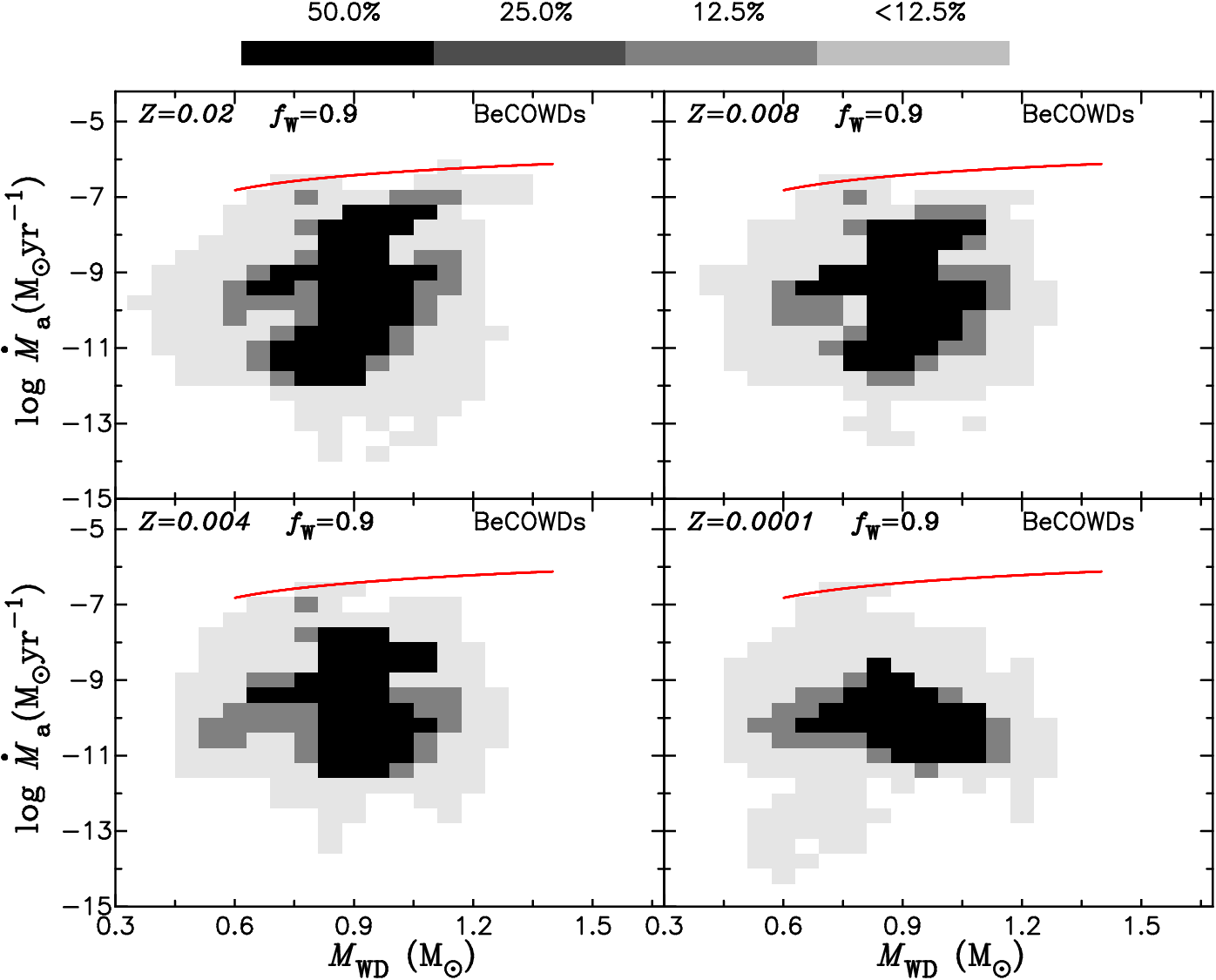}&
\includegraphics[totalheight=3.2in,width=3.5in,angle=-90]{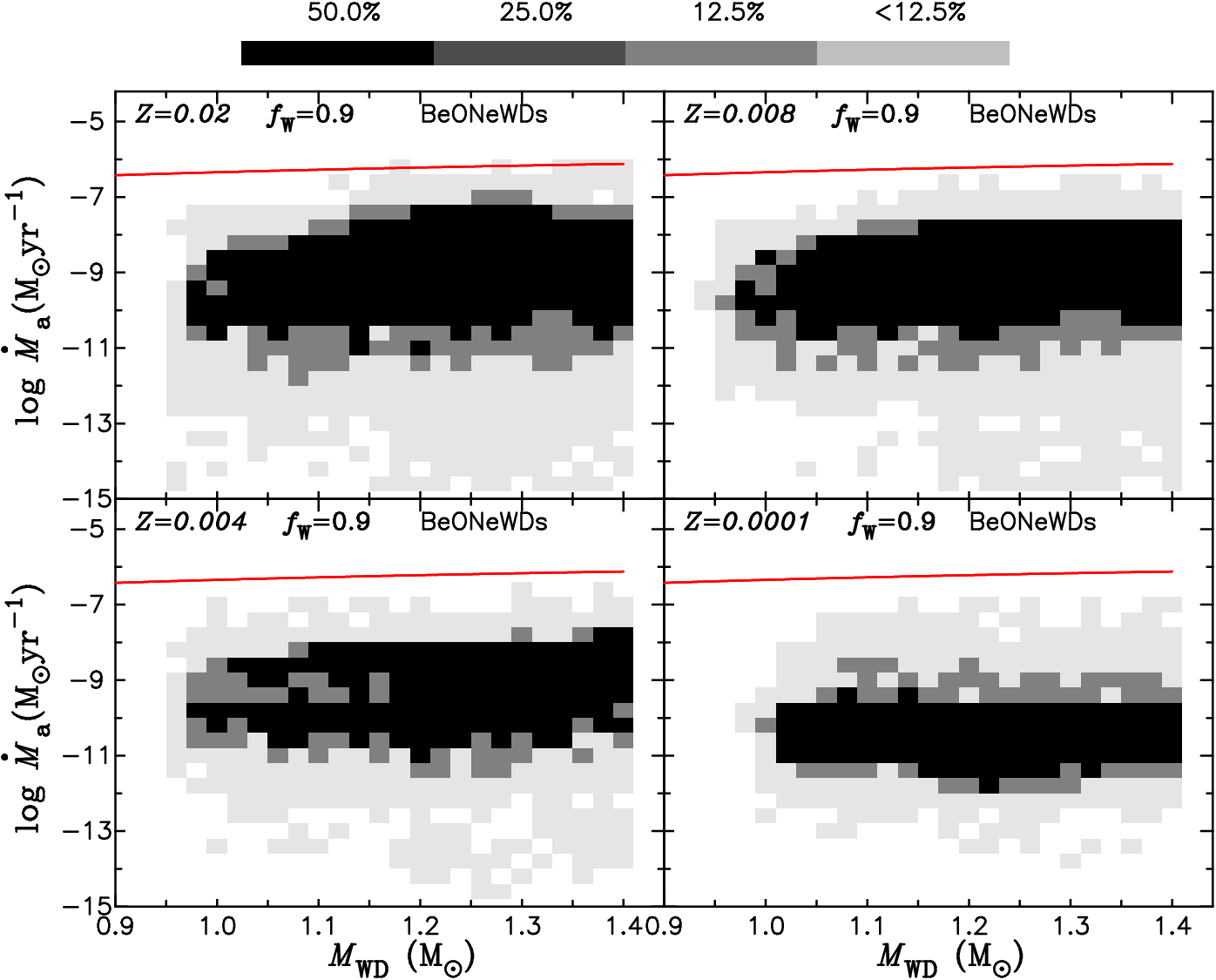}\\
\end{tabular}
\caption{Distributions of WD mass vs. the mass-accretion rates in BeWDs. Red lines represent the critical mass-accretion rate ($\dot{M}_{\rm cr}$)
for the accreted hydrogen burning steadily on the surface of the WD. }
\label{fig:snia}
\end{figure*}

Figure \ref{fig:snia} shows the distribution of WD'mass and mass-accretion rate in BeWDs.
Obviously, the mass-accretion rates of WDs in BeWDs almost are lower than $\dot{M}_{\rm cr}$,
and the majority of them are lower than $\frac{1}{8}\dot{M}_{\rm cr}$.
Therefore, in our simulations, there is not a sample in which COWD's mass can reach up to 1.35 M$_\odot$
and ONeWD's mass can reach up to 1.44 M$_\odot$ in BeWDs.
Therefore, BeWDs hardly occur SN Ia.

\section{Conclusions}
Using the method of population synthesis, we investigate
the formation and the destiny of BeWDs.
The effects of the metallicity and the combining parameters
$\lambda\times\alpha_{\rm CE}$ of CEE on BeWD population are discussed.
For $\lambda\times\alpha_{\rm CE}=0.5$, about 3.7\% ($Z=0.0001$)-4.1\% ($Z=0.02$) of binary systems can evolve into BeWDs;
about 60\% ($Z=0.0001$)-70\% ($Z=0.02$) of BeWDs include a COWD, and 30\%-40\% of them
have an ONeWD;
about 40\% ($Z=0.0001$) -45\% ($Z=0.02$) of BeCOWDs have undergone CEE,
about 35\% ($Z=0.0001$) -50\% ($Z=0.02$) of them have experienced heavy mass transfer, and
about 10\% ($Z=0.0001$) -20\% ($Z=0.02$) of them exchange materials via stellar winds;
for BeONeWDs, the above proportions are 50\%-60\%, 20\%-30\% and 20\%-30\%,
respectively. Changing the combining parameter $\lambda\times\alpha_{\rm CE}$ from 0.25 to 0.5, it introduces
an uncertainty with a factor of about 1.3 on BeWD populations.
It mainly affects BeWDs which are formed via CEE of the binary systems with $10^2<P_{\rm orb}^{\rm i}<\sim 10^3$  days.

About 30\%-50\% of BeWDs come from BesdOBs which are important
progenitors. Our results can cover well the observational properties of BesdOBs' population,
including rare sources: HR 6819 and ALS 8775. BesdOBs mainly evolve into the BeWDs with orbital periods
shorter than about 300 days. About 60\%-70\% of BeWDs occur a merger between a WD and a non-degenerated star
in which 90\% are H-rich stars and the left are He stars.
About 30\%-40\% of BeWDs turn into double WDs which are potential GWs of LISA mission
at a frequency band between about $3\times10^{-3}$ and $\times10^{-2}$ Hz.
Due to a low mass-accretion rate of WD, BeWDs hardly become the progenitors of SN Ia.

One should note that the uncertainties of BPS in our work only result from the metallicity and the CEE.
It is well known that many input parameters (accretion efficiency during mass transfer,
the criteria of dynamical mass transfer, et al.) can affect the results of BPS.
If the effects of the combining parameters are considered, the uncertainties of BPS should be larger.

\begin{acknowledgements}
This work received the generous support of  the Natural Science Foundation of Xinjiang No.2021D01C075,
the National Natural Science Foundation of China, project Nos. 12163005, U2031204 and 11863005,
the science research grants from the China Manned Space Project with NO. CMS-CSST-2021-A10.
\end{acknowledgements}
\bibpunct{(}{)}{;}{a}{}{,}

\end{document}